\pdfoutput=1
\documentclass[11pt,twoside,openany]{book}

\usepackage[letterpaper,lmargin=3cm,rmargin=2cm,bmargin=3.1cm,tmargin=3.1cm]{geometry} 

\usepackage[bookmarksopen,colorlinks]{hyperref}  
\hypersetup{citecolor=black, filecolor=black, linkcolor=black, urlcolor=black} 

\usepackage{graphicx}  
\usepackage{tikz}
\usepackage{flafter}  
\usepackage[list=off]{caption}
\usepackage{subcaption}
\usepackage{cite}
\usepackage{booktabs}

\usepackage{amsthm}
\usepackage{amsmath,amssymb}  
\usepackage{bm}  
\usepackage[utf8]{inputenc} 
\usepackage{textcomp} 
\usepackage[spanish]{babel}
\usepackage[plain]{fancyref}

\usepackage{enumitem}
\setlist{itemsep=.5pt}

\usepackage{fancyhdr} 
\usepackage{aas_macros} 
\usepackage[]{tocbibind} 

\makeatletter 
\DeclareRobustCommand*{\bfseries}{%
  \not@math@alphabet\bfseries\mathbf
  \fontseries\bfdefault\selectfont
  \boldmath
}
\makeatother
    
\newcommand*{\mychemistry}[2]{$^{\text{#2}}\text{#1}$}
\newcommand*{\mychemistryg}[3]{$^{\text{#2}\!}_{\text{#3}\!}\text{#1}$}

\begin{document}

    \renewcommand{\tablename}{Tabla}

    \renewcommand{\thefootnote}{\roman{footnote}}

    \lefthyphenmin=2
    \righthyphenmin=2

    \frontmatter

\pagestyle{empty}
\begin{titlepage}
\begin{center}

\vfill

\Large

Universidad de La Habana 

\medskip

Facultad de Física

\vspace{10mm}

\centerline{\includegraphics[height=30mm, keepaspectratio]{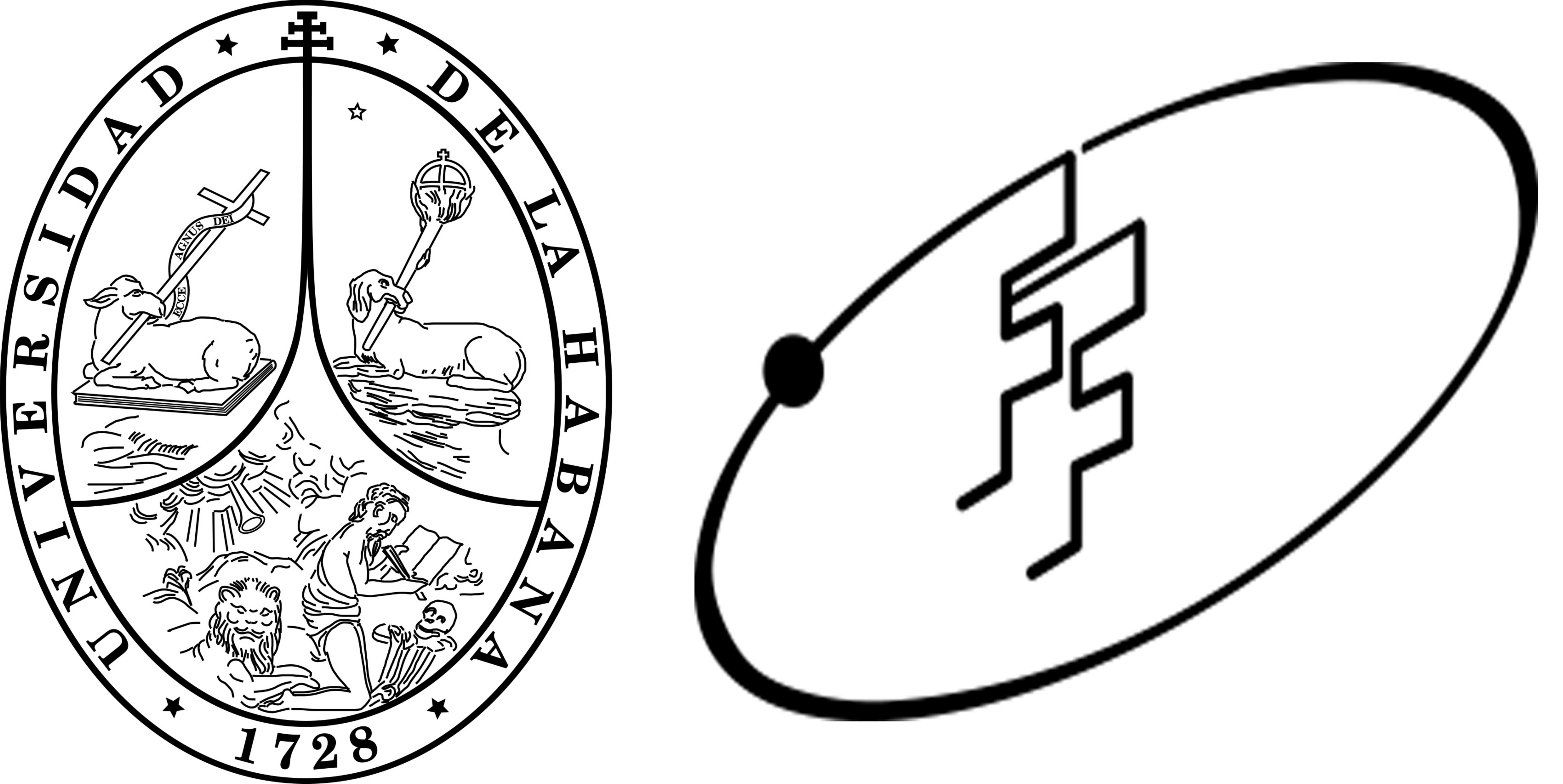}}

\vfill

{\bf\Large TESIS DE DIPLOMA}

\vspace{7mm}

{\LARGE\bfseries Enanas Blancas Magnetizadas}

\vfill

\begin{tabular}{rl}

\textbf{Autor:} & Diana Alvear Terrero \\
\noalign{\vspace{10mm}}

\textbf{Tutores:} & Dra. Aurora Pérez Martínez, ICIMAF \\
			      & Dr. Daryel Manreza Paret, Facultad de Física \\
\noalign{\vspace{2mm}}
\end{tabular}

\vfill

\Large
La Habana, 2015

\end{center}
\end{titlepage}
\cleardoublepage
    
\indent
\vfill

\begin{quote}
A abuela Niurka, que me ayudó a escribir estas líneas, y al resto de mi gran familia, por su apoyo incondicional.

A mis tutores,  Aurorita y Daryel, que me introdujeron en la física de los Objetos Compactos. A Hugo Pérez y el resto de la familia del ICIMAF. Gracias por  su acogida, su dedicación y su ejemplo.

A mis compañeros de grupo y amigos, por la ayuda y las ideas que intercambiamos siempre. 

A quienes de una forma u otra han hecho posible llevar a buen término este trabajo.

\end{quote}

\vfill
\cleardoublepage

\pagestyle{plain}
    \pagestyle{empty}


\vfill

\begin{center}
\section*{Resumen}
\addcontentsline{toc}{chapter}{Resumen}
\end{center}
\vspace{24pt}

\begin{quote}
El propósito de esta tesis es obtener ecuaciones de estado más realistas para describir la materia que forma a las enanas blancas magnetizadas, y resolver con ellas las ecuaciones de estructura de estos objetos compactos.

Las ecuaciones de estado se determinan teniendo en cuenta la aproximación de campo magnético débil $B< B_c$ ($B_c=4.41\times10^{13}$ G) para el gas de electrones de la estrella.
El campo magnético, aún para los valores moderados presentes en las enanas blancas, introduce presiones anisotrópicas. Además, consideramos la corrección a la energía y la presión debida a la interacción Coulombiana del gas de electrones con los iones localizados en una red cristalina.
Al introducir esta corrección, disminuye la energía y la presión del sistema, efecto que se magnifica al emplear elementos químicos más pesados.

Por otra parte, se resuelven las ecuaciones de estructura de Tolman-Oppenheimer-Volkoff en simetría esférica de manera independiente para la presión perpendicular y la paralela, lo cual confirma que es necesario utilizar ecuaciones de estructura acordes a la axisimetría del sistema anisotrópico. Por tanto, estudiamos las soluciones de  las ecuaciones de estructura en coordenadas cilíndricas. En este caso, se obtiene la masa por unidad de longitud en vez de la masa total de la enana blanca.

\end{quote}

\vfill

\cleardoublepage

    \tableofcontents
    
    
    \mainmatter

    \pagestyle{fancy}

    
\chapter*{Introducción} \label{intro}
\addcontentsline{toc}{chapter}{Introducción}

A medida que una estrella brilla, pierde, en un tiempo finito, su reserva de energía nuclear. Cuando la ha agotado toda, sobreviene la llamada muerte estelar. En este momento, la presión del gas caliente en el interior no puede soportar el peso de la estrella y esta colapsa. El remanente estelar se conoce como objeto compacto (OC), y en dependencia de la masa de la progenitora, será una enanas blanca (EB), una estrella de neutrones (EN) o un agujero negro (AN) . 

Nuestra galaxia está poblada por miles de millones de enanas blancas, unos pocos cientos de millones de estrellas de neutrones y probablemente, cientos de miles de agujeros negros. De todos estos objetos, sólo una pequeña fracción se ha detectado hasta el momento:
 miles de enanas blancas, alrededor de 2 000 estrellas de neutrones y solo unas pocas docenas de agujeros negros \cite{2005NJPh....7..199N,2006msu..conf..145C}. 

Los objetos compactos tienen densidades extremadamente altas, de $10^7$ g\hspace{2pt}cm$^{-3}$ para las enanas blancas y $10^{15}$ g\hspace{2pt}cm$^{-3}$ en el caso de las estrellas de neutrones\footnote{La densidad media de la Tierra es $5.514$ g\hspace{2pt}cm$^{-3}$ y la densidad nuclear es $2.04\times10^{14}$ g\hspace{2pt}cm$^{-3}$.}, y son laboratorios naturales que no se pueden describir obviando alguna de las cuatro fuerzas de la naturaleza: la débil, la fuerte, la electromagnética y la gravitacional. Tales condiciones extremas no se han podido obtener en laboratorios terrestres. Luego, solo disponemos de observables astrofísicos para inferir las propiedades de objetos tan densos y masivos.

La primera enana blanca observada fue la compañera de Sirio A (Sirio en aquel entonces), cuando en 1844, el astrónomo Friedrich Bessel notó un ligero movimiento de vaivén, como si a su alrededor orbitara un objeto no visto. En 1863, Alvan Clark observó finalmente dicho objeto, que se identificó como una enana blanca, llamada Sirio B.

Para los físicos y astrónomos  de principios del siglo XIX resultaba un desafío explicar qué fuerza estaba compensando la atracción gravitacional en un astro como Sirio B, tan pequeño y poco luminoso, pero tan masivo, ya que en un objeto así han cesado las reacciones de fusión, por lo que la presión térmica es muy débil y no puede contrarrestar la fuerza gravitacional.

La explicación a este enigma aparece con las teorías cuánticas y la formulación por Paul Dirac en 1926 de la estadística a la que obedecen los fermiones, bautizada posteriormente como \emph{estadística de Fermi-Dirac}. Esta tiene en cuenta el Principio de exclusión de Pauli: dos fermiones (partículas idénticas e indistinguibles) no pueden ocupar el mismo estado cuántico. De este modo, en diciembre de 1926 R. H. Fowler \cite{1926MNRAS..87..114F} obtuvo que un gas de fermiones  denso a temperatura cero es capaz de ejercer una presión no nula que puede entonces sostener el colapso.

En 1925, el espectro de Sirio B confirmó que es una estrella con aproximadamente la misma temperatura de Sirio A. Ambas están en órbita, una alrededor de la otra, y constituyen lo que se llama un sistema binario, lo cual permitió que se determinaran sus masas usando la tercera Ley de Kepler, resultando masas de $2.3 M_{\odot}$ y $1M_{\odot}$\footnote{$M_\odot$ es la masa de Sol (ver \Fref{tab:Sol}).} para Sirio A y B, respectivamente. 
También se obtuvo un diámetro de 10 000 km para Sirio B, mientras que el de Sirio A es 1 000 000 km. Estas mediciones se refinaron con la ayuda del telescopio espacial Hubble en el 2005, siendo el diámetro obtenido para Sirio B de 12 000 km \footnote{Comparable con el diámetro de la Tierra: 12 742 km.} y la masa de $0.98 M_{\odot}$, en  concordancia con los resultados previos.

Por otra parte, el descubrimiento del neutrón por James Chadwick en 1932\cite{1932Natur.129Q.312C}, condujo a Walter Baade y Fritz Zwicky a proponer la existencia de ENs en 1934 \cite{1934PNAS...20..254B}. Esta idea alcanzó relevancia en 1967, con el  azaroso descubrimiento por Jocelyn Bell y Antony Hewish de la primera fuente de radio (Pulsar, del inglés \emph{pulsating star}), en el Radio Observatorio de Cambridge~\cite{1968Natur.217..709H}. 

Las
enormes densidades concentradas en espacios muy pequeños de los objetos compactos, hacen  que para estudiar su equilibrio hidrodinámico y describirlos se requiera la Teoría General de la Relatividad (TGR), tomando la presión del gas degenerado de electrones/neutrones como la que compensa la presión gravitacional. 
La aplicación de la TGR, conduce a la existencia de valores máximos para las masas de estos objetos. 

En las EBs  el límite m\'aximo para la masa es de $1.44 M_{\odot}$, y se conoce como masa de Chandrasekhar \cite{1931ApJ....74...81C}. Más all\'a de este valor, la presi\'on degenerada del gas de electrones no puede compensar la presi\'on gravitacional. Además, tienen radios $R\lesssim1\,000$ km  y densidades de $(10^7-10^8)$ g\hspace{2pt}cm$^{-3}$.  Estas características, junto con la baja luminosidad, hacen que las enanas blancas  sean difíciles de detectar y que se ubiquen a la izquierda de la secuencia principal en el Diagrama de Hertzsprung-Russell (ver Sección~\ref{cap11}). 
Por otra parte, muchas EBs se localizan en sistemas binarios, en los que se deposita material de la estrella compañera en un disco de acreción alrededor de la enana blanca, lo cual da origen a las Supernovas Tipo Ia. 

Asimismo, para las ENs se encuentran soluciones de masas entre $1.44M_{\odot}$ y $2M_{\odot}$, radios entre $10$ km y $20$ km \cite{1939PhRv...55..374O} y densidades que varían en el rango de $(10^7-10^{15})$ g\hspace{2pt}cm$^{-3}$, siendo los objetos más densos en el universo. No son visibles con telescopios ópticos y por tanto no podemos representarlas en el Diagrama de Hertzsprung-Russell (ver Sección~\ref{cap11}).

En los últimos 20 años se ha recopilado una gran cantidad de datos observacionales  de los objetos compactos gracias a la construcción de radiotelescopios, el desarrollo de los satélites y de instrumentos de detección de radiación  X y  gamma para colocarlos en ellos. 

En la Tierra podemos mencionar una serie de radiotelescopios, el VLA (\emph{Very Large Array}) en Estados Unidos, el RTGM (Radio Telescopio Gigante en Metro-ondas) en la India, el ATCA (\emph{Australia Telescope}), el ALMA (\emph{Atacama Large Milimeter Array}), y el SKA (\emph{Square Kilometer Array}). Destacan también, los experimentos de sondeo del espacio como el \emph{Hamburg/ESO Quasar Survey},
el \emph{Edinburgh-Cape survey} 
y el \emph{Sloan Digital Sky Survey} (SDSS).

El observatorio en satélite más conocido es el Telescopio Espacial Hubble (TEH), puesto en \'orbita el 24 de abril de 1990 como un proyecto conjunto de la NASA (\emph{National Aeronautics and Space Administration}) y de la ESA (\emph{European Space Agency}), que puede captar radiación en el visible, en el infrarrojo próximo y en el ultravioleta. El TEH  ha tomado, por ejemplo, numerosos datos de supernovas, enanas blancas y estrellas de neutrones. Su papel ha sido decisivo en la “cacería” de Supernovas tipo Ia. 

Actualmente se dispone además de un conjunto de observatorios como el Observatorio de Rayos X Chandra y el Observatorio de Rayos Gamma Compton, que se encuentran en órbita al igual que el Hubble. Ellos están en  una posición clave para obtener  nuevas imágenes de alta resolución espectral y estudiar objetos compactos en sistemas aislados, binarios y en las regiones de galaxias que muestran actividad energética inusual (AGN, en inglés \emph{Active galactic nucleus}).

La presencia de elevados campos magnéticos caracteriza  también a los objetos compactos. Las observaciones permiten inferir los campos magnéticos superficiales, pero no son capaces de estimar los  valores en el interior. A las EBs se les atribuye campos magn\'eticos entre $10^3$\,G y $10^{9}$\,G \cite{2013MNRAS.430...50C} \footnote{El campo magnético  terrestre es de $0.5$ G.} en la superficie; en tanto para las ENs, se asocian campos magnéticos superficiales de $10^{12}$ G a los llamados radio pulsares y de $10^{15}$ G a las nombradas Magnetars (\emph{Magnetic Stars}-Estrellas Magnéticas)~\cite{1983bhwd.book.....S}.  

Los valores de hasta $10^{12}$~G pueden ser  explicados por la amplificación del campo magnético de las estrellas progenitoras después de la explosión de  la Supernova~\cite{1983bhwd.book.....S}. Sin embargo, intensidades mayores, en particular los campos magnéticos asociados a las Magnetars, precisan de modelos más elaborados que aún est\'an en discusi\'on \cite{1996AIPC..366..111D}. Uno de ellos es el efecto dinamo: basado en la circulaci\'on del gas por convección dentro de la estrella.

Los campos magnéticos intensos modifican el comportamiento de la materia a escala microscópica y con ello la estructura estelar. Un estimado teórico de los valores m\'aximos de los campos magnéticos que podr\'{i}an  llegar a tener estos objetos en el interior puede darse a trav\'es del teorema escalar del Virial \cite{Shapiro1991ApJ, 1978vtsa.book.....C}.
Esto significa equiparar la energ\'{\i}a gravitacional con la magnética. 
Para EBs encontramos que los campos magnéticos estimados para masas de $M = 1.1 M_\odot$
y radios de $R = 0.02R_\odot$, son como máximo de aproximadamente $10^{12}$ G; en tanto para ENs encontramos que si la masa es $M = 1.44 M_\odot$  y el radio $R = 10^6$ cm$=10^{-4}R_\odot$, se obtiene un campo magnético m\'aximo del orden de $10^{18}$ G.

Recientemente, han sido reportadas observaciones de supernovas tipo Ia SN~2006gz, SN~2007if, SN~2009dc, SN~2003fg con valores de luminosidad muy elevados y baja energ\'{\i}a cin\'etica \cite{2010ApJ...713.1073S}. Estas supernovas han tratado de ser explicadas~\cite{2013PhRvL.110g1102D,2012IJMPD..2142001D,2014JCAP...06..050D} a partir EBs progenitoras con masas mayores que el límite de Chandrasekhar impuesto por la teoría.

En particular, se ha asociado la presencia de campos magn\'eticos fuertes ($10^{14}-10^{15}$) G en las EBs, como la causa de que lleguen a tener masas mayores que las de Chandrasekhar \cite{2013PhRvL.110g1102D}. Teniendo en cuenta que estos modelos  fueron construidos ignorando aspectos microfísicos y macrofísicos esenciales, en \cite{2014ApJ...794...86C} y \cite{Paret:2015RAA} han sido descartados. 

La motivación de esta tesis parte del intento de estudiar el papel que juega la presencia de campos magn\'eticos en las enanas blancas y construir modelos más realistas, tanto para las ecuaciones de estado como para las ecuaciones de estructura de las EBs magnetizadas.

Para ello, nos hemos propuesto:
\begin{itemize}
\item Estudiar una ecuación de estado que describa de manera más realista una EB magnetizada, considerando no solo el papel del campo magnético, sino también la presencia de la interacción de los electrones y los iones, que forman una red cristalina. Esto presupone considerar el plasma de electrones e iones de la estrella en presencia de campo magnético.
Usaremos el límite de campo débil $B< B_c$ ($B_c=4.41\times10^{13}$ G), ya que los campos magnéticos inferidos para EBs son de hasta $10^{9}$ G en la superficie y $10^{12}$ G en el interior. 

\item Resolver las ecuaciones de estructura de dichos objetos en el marco de la Teoría General de la Relatividad, lo cual nos permite obtener los observables macroscópicos: masas y radios.  Analizaremos las soluciones de las ecuaciones de estructura para simetría tanto esférica como cilíndrica. Esta última es más realista debido a la anisotropía en las presiones que introduce el campo magnético.
\end{itemize}

La tesis complementa y le da continuidad al trabajo \cite{phdDaryel}, donde se estudió tanto las ecuaciones de estado como las de estructura para una EB magnetizada. El gas  magnetizado degenerado de electrones en \cite{phdDaryel} fue considerado libre y no se tomaron en cuenta las interacciones con los iones. Partiremos tal y como se hizo en \cite{phdDaryel},  de suponer que el campo magn\'etico que existe en la estrella es dipolar y  constante. No pretendemos abordar los mecanismos que lo producen ni sus orígenes.

La validez de la aproximaci\'on de campo magnético constante se debe a que, aunque la intensidad del campo magnético var\'{\i}a algunos \'ordenes desde el n\'ucleo de la estrella a la superficie, la escala de variaci\'on del mismo dentro de la estrella es mucho mayor que la variaci\'on de la escala del campo magn\'etico microsc\'opico 
 \cite{2000ApJ...537..351B}. Es decir, la escala microsc\'opica magn\'etica est\'a dada por la longitud magnética $l_m \sim 1/eB$ y se satisface ampliamente que $R\gg l_m$, donde $R$ es el radio de la estrella y tiene un valor aproximado de 1 000 km (20 km), para EBs (ENs).

Hemos organizado la tesis en cuatro capítulos, dos apéndices y finalmente, las conclusiones y recomendaciones y la bibliografía. Los resultados originales de la tesis y las contribuciones del autor a la temática se incluyen en los Capítulos \ref{cap3} y \ref{cap4}. 
\begin{itemize}
\item El Capítulo \ref{cap1}, a modo introductorio, presenta las características fundamentales de la enanas blancas y las enanas blancas magnetizadas.
\item En el Capítulo \ref{cap2},  partiendo del tensor de energía-momento de un gas degenerado de electrones en ausencia y presencia de campo magnético, se obtienen las propiedades termodinámicas correspondientes.
\item En el Capítulo \ref{cap3}, se analizan las ecuaciones de estado anisotrópicas para enanas blancas no magnéticas y enanas blancas magnetizadas en el límite de campo débil. Se considera además la inclusión de la interacción de los electrones con la red cristalina y se consideran distintas composiciones químicas. 
\item En el Capítulo \ref{cap4}, se utiliza las ecuaciones de estado previamente obtenidas para resolver las ecuaciones de estructura en simetría esférica, discutiendo el impacto de la anisotropía de las presiones  en las relaciones masa-radio. Además, se resuelven las ecuaciones de estructura en simetría cilíndrica. 
\end{itemize}


\chapter{Enanas blancas}
\label{cap1}

El estudio de la estructura de las enanas blancas entraña la creación de modelos en los que se tiene en cuenta las propiedades microfísicas de las mismas para construir ecuaciones de estado correspondientes.  Una vez obtenidas las ecuaciones de estado, se resuelven las ecuaciones de estructura, cuya solución devuelve distintas configuraciones para los observables macroscópicos.

En este capítulo discutiremos de forma sintetizada el proceso de evolución estelar para estrellas de masa pequeña, así como las características principales de las enanas blancas y las enanas blancas magnéticas.

\section{Evolución estelar} \label{cap11}

De manera general, la evolución estelar puede describirse como una lucha entre la fuerza gravitatoria, que desde la formación de una estrella a partir de una nebulosa, tiende a comprimirla; y la fuerza nuclear proveniente de las reacciones de fusión, fisión y desintegración radiactiva, que se opone a la contracción. Eventualmente el combustible nuclear se agota, y la estrella colapsa gravitacionalmente. Cómo se desarrolle específicamente este proceso, depende de las características de la estrella, particularmente de la masa y la composición.

Los elementos químicos presentes en una estrella determinan las líneas de absorción observadas en los espectros estelares, que se disponen en una secuencia continua según la intensidad de las líneas. 
A modo de clasificación, se designan los tipos O, B, A, F, G, K, L, M, y T, que conforman el llamado sistema de Harvard, desarrollado en el Observatorio de Harvard a inicios del siglo XX. 

Inicialmente, dicha clasificación cubría desde las estrellas azules más calientes de tipo O hasta las más frías de clase M, pero posteriormente se le han ido añadiendo otras letras, como la T para las infrarrojas. Además, se utilizan los subíndices del 0 al 9 para indicar las sucesiones dentro de cada clase, donde A0 especifica las estrellas más calientes de la clase A, mientras que A9 se refiere a las más frías. 

A partir del tipo espectral, la temperatura, y la ley de Stefan-Boltzmann $R \sim \sqrt{L}/T^2$ (siendo $R$ el radio de la estrella, $L$ su luminosidad y $T$ la temperatura efectiva), vemos que dada una temperatura, las estrellas más luminosas tienen mayores dimensiones.  

\begin{figure}[ht]
\centering
\includegraphics[width=.65\textwidth]{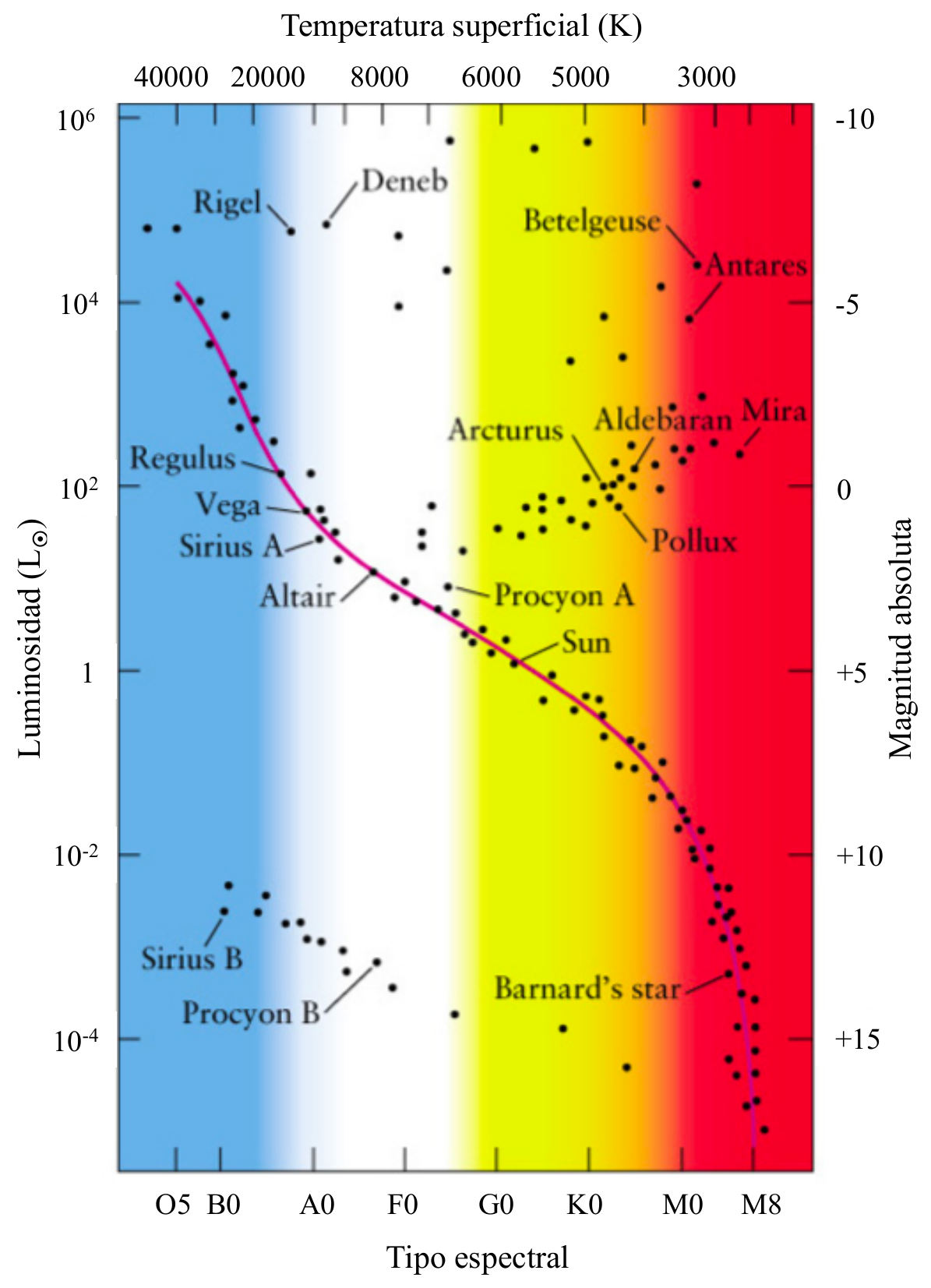}  
\caption{Diagrama H-R.}
\label{fig:HR}
\end{figure}

Esta relación se muestra en el Diagrama de Hertzsprung-Russell o Diagrama H-R (Figura \ref{fig:HR}), que fue desarrollado de manera independiente a inicios del siglo XX por Ejnar Hertzsprung y Henry Norris Russell. El gráfico original de Hertzsprung mostraba la luminosidad de las estrellas en función de su color, en tanto el de Russel representaba la luminosidad contra la clase espectral.

En la Figura \ref{fig:HR} se observa que las estrellas se agrupan en regiones definidas, lo cual permite localizar, por ejemplo, las gigantes y las enanas blancas. La diagonal que va desde el extremo superior izquierdo hasta el inferior derecho se conoce como secuencia principal. La relación entre la luminosidad y el brillo en la secuencia principal indica que la posición de cada estrella depende de su masa.

En particular, el remanente luego de la llamada muerte estelar, está condicionado por la masa inicial de la estrella, como aparece en la Tabla \ref{tab:dist_masa}. Los valores dependen de las observaciones y, aunque en general concuerdan con los tabulados, varían de un autor a otro.

\begin{table}[h]
\centering
\begin{tabular}{cc} \toprule
Rango de masa aproximado & Resultado esperado \\  \midrule
$0.08 M_{\odot} \lesssim M \lesssim 10 M_{\odot}$  	 & Enana blanca y nebulosa planetaria.\\
$10 M_{\odot} \lesssim M \lesssim 25 M_{\odot}$&  Estrella de neutrones y supernova. \\
 $25 M_{\odot}\lesssim M \lesssim 60 M_{\odot}$ & Agujero negro y supernova o explosión de rayos gamma.\\ \bottomrule
 \end{tabular}
\caption{Posibles destinos de las estrellas estables como función de la masa \cite{1983bhwd.book.....S}.}
\label{tab:dist_masa}
\end{table}

\begin{figure}[h]
\centering
\includegraphics[width=.6\textwidth]{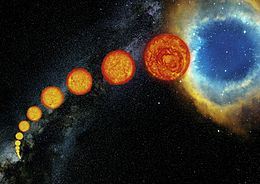}
\caption{Representación de la evolución de estrellas como el Sol.}
\label{fig:sunlikestars}
\end{figure}

El ciclo de vida de las estrellas de masa pequeña como nuestro Sol (Figura \ref{fig:sunlikestars}) comienza fusionando hidrógeno. Cuando este elemento se agota, el núcleo se contrae sobre su propio peso y las capas externas se expanden formándose una gigante roja. Eventualmente, la temperatura del núcleo es suficientemente alta para que se inicien las reacciones de fusión del helio.  Esto hace que la estrella se torne inestable, y causa que se eyecten las capas exteriores formando una nebulosa planetaria. El núcleo remanente es una enana blanca. 

Para estrellas de masas intermedias ($3M_\odot<M<9M_\odot$), que terminan como enanas blancas de carbono y oxígeno, el proceso evolutivo es similar.

Las estrellas pueden encontrarse formando sistemas binarios, y en el caso de que uno de los componentes sea un OC, este puede interactuar con la otra estrella a través de un disco de acreción de masa. En un sistema binario con una EB, la acreción de masa por parte de esta puede llevarla a alcanzar la masa de Chandrasekhar. Los electrones y protones comienzan a reaccionar generando neutrones (neutronización), y la estrella colapsa produciendo una Supernova tipo Ia. 

Este tipo de supernovas son muy importantes en las mediciones de distancias en el universo porque poseen una cantidad estándar de combustible y un mecanismo de explosión común (candelas estándar). Así, con mediciones de distancias basadas en las Supernovas tipo Ia se ha inferido la aceleración de la expansión del universo \cite{1998AJ....116.1009R} y la presencia de energía oscura \cite{1999ApJ...517..565P}.

Por otra parte, las EBs sin fuentes de energía termonuclear se enfrían con el tiempo, a medida que radían su energía residual. Eventualmente, se convierten en enanas negras, las cuales cristalizarán al final según la teoría. Como contienen predominantemente carbono, son nombradas “diamantes del Universo”.

\section{Características}

Las enanas blancas tienen típicamente masas medias de $0.663\, M_\odot$ \cite{2013A&A...559A.104T}, siendo las más pequeñas del orden de $0.17\, M_\odot$ \cite{2007ApJ...660.1451K}, y las mayores aproximadamente $1.33\, M_\odot$ \cite{2007MNRAS.375.1315K}. Los radios característicos se estiman entre los $0.008\,R_\odot$ y $0.02\,R_\odot$, lo cual es comparable con el radio de la Tierra $R_T \sim 0.009\,R_\odot$; y las densidades en el interior varían entre $10^4$ g cm$^{-3}$ y $10^9$ g cm$^{-3}$, con densidad media \mbox{$10^6$ g cm$^{-3}$}. Las temperaturas superficiales van desde 5 000 hasta 80 000 K.

Están formadas por una atmósfera de radio muy pequeño, por lo que puede despreciarse en los modelos; y un núcleo fundamentalmente de He, C y O, aunque pueden encontrarse elementos más pesados por efectos de acreción de masa, ya sea del medio interestelar o de una compañera. Su clasificación espectral se simboliza por la letra D (proveniente de \emph{dwarf}, enana en inglés), seguida de una de las letras A, B, C, O, Z, Q, en dependencia de la composición de su atmósfera:

\begin{itemize}
  \item[] DA: atmósfera rica en hidrógeno, con líneas fuertes de HI (constituyen el 80 \% de las EBs).
  \item[] DB: atmósfera rica en helio, con líneas fuertes de HeI.
  \item[] DC: espectro continuo.
  \item[] DO: atmósfera rica en helio, con líneas fuertes de HeII.
  \item[] DZ: líneas fuertes  de metales (por ejemplo CaI, MgI, FeI, excluyendo al carbono).
  \item[] DQ: líneas fuertes de carbono.
\end{itemize}
\begin{figure}[h]
\centering
\includegraphics[width=.9\textwidth]{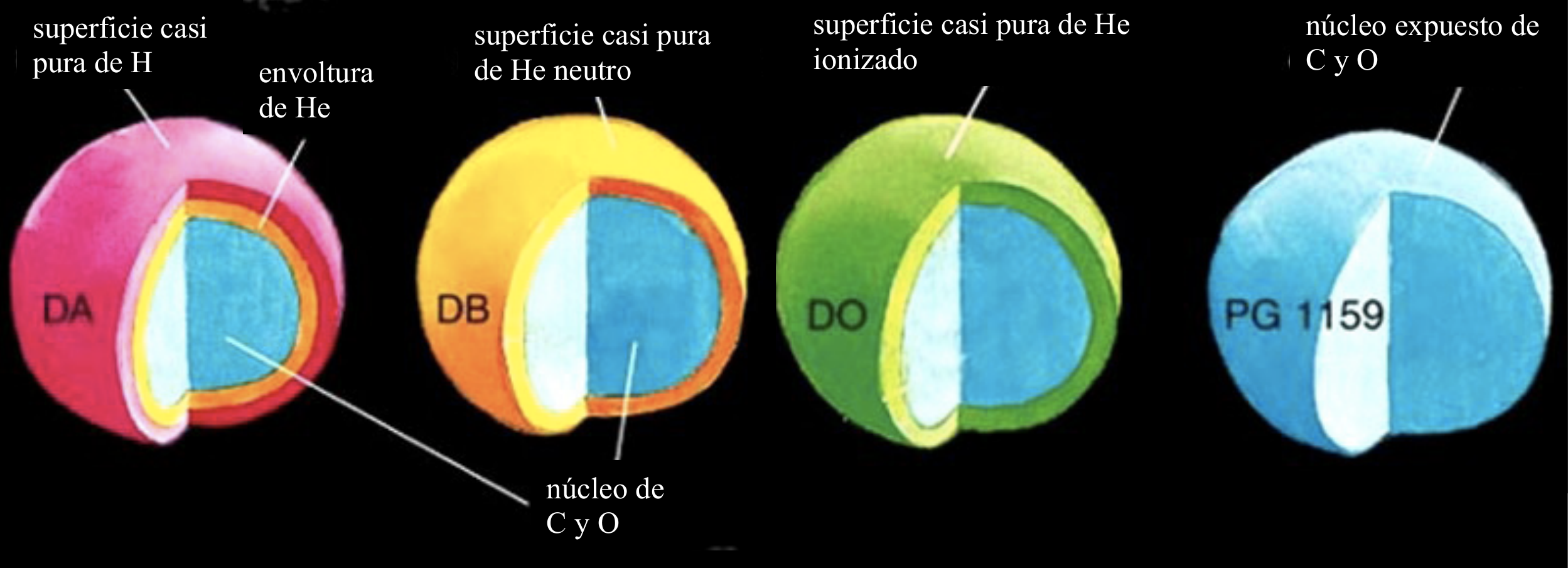}
\caption{Composición y estructura de algunas clases de enanas blancas.}
\label{fig:ceEB}
\end{figure}

\section{Campos magnéticos en las enanas blancas}

En 1970, Kemp \cite{1970ApJ...161L..77K} demostró que la radiación proveniente de Grw$+70^\circ8247$ \cite{1934PASP...46..287K} presentaba una fuerte polarización circular. Posteriormente, las líneas espectrales observadas fueron identificadas como líneas de hidrógeno desplazadas por efecto Zeeman en un campo magnético de aproximadamente $(10^8-3.2\times10^8)$ G \cite{1985ApJ...292..260A,1985ApJ...289L..25G,1988ApJ...327..222W}, por lo que fue clasificada como una enana blanca magnética (EBM). 

Actualmente, el número de enanas blancas magnéticas aisladas descubiertas asciende a aproximadamente $250$ con el campo magnético bien determinado y más de $600$ si se cuenta los objetos donde el campo no se ha obtenido \cite[y referencias encontradas en él]{2015SSRv..191..111F}. 

También, existen sistemas binarios llamados variables cataclísmicas magnéticas (VCMs) compuestos por una EBM y una estrella de la parte inferior de la secuencia principal muy próximas entre sí, en los que la enana blanca presenta un campo magnético lo suficientemente fuerte como para  afectar el disco de acreción. Estos sistemas se dividen en dos grupos, en dependencia de la intensidad del campo magnético: polares (estrellas de tipo AM Her) y polares intermedios (estrellas de tipo DQ Her). 

Los polares se caracterizan porque el momento magnético de la EBM, del orden de \mbox{$10^{33}$ G cm$^3$} o mayor, es suficiente para que las estrellas roten de forma sincronizada con período orbital $(70-480)$ min.  En cambio, el momento magnético de la EBM en los polares intermedios es menor que en el caso de los polares, y no logra sincronizar la órbita de la EB con el período orbital del sistema. Hoy día, el número de VCMs listadas es de aproximadamente 170, aunque solo se tiene la intensidad del campo magnético para la mitad de ellas, la mayor parte en un rango de $(7\times10^6-2.3\times10^8)$ G.

\begin{figure}[h]
\begin{center}
	\includegraphics[width=.55\textwidth]{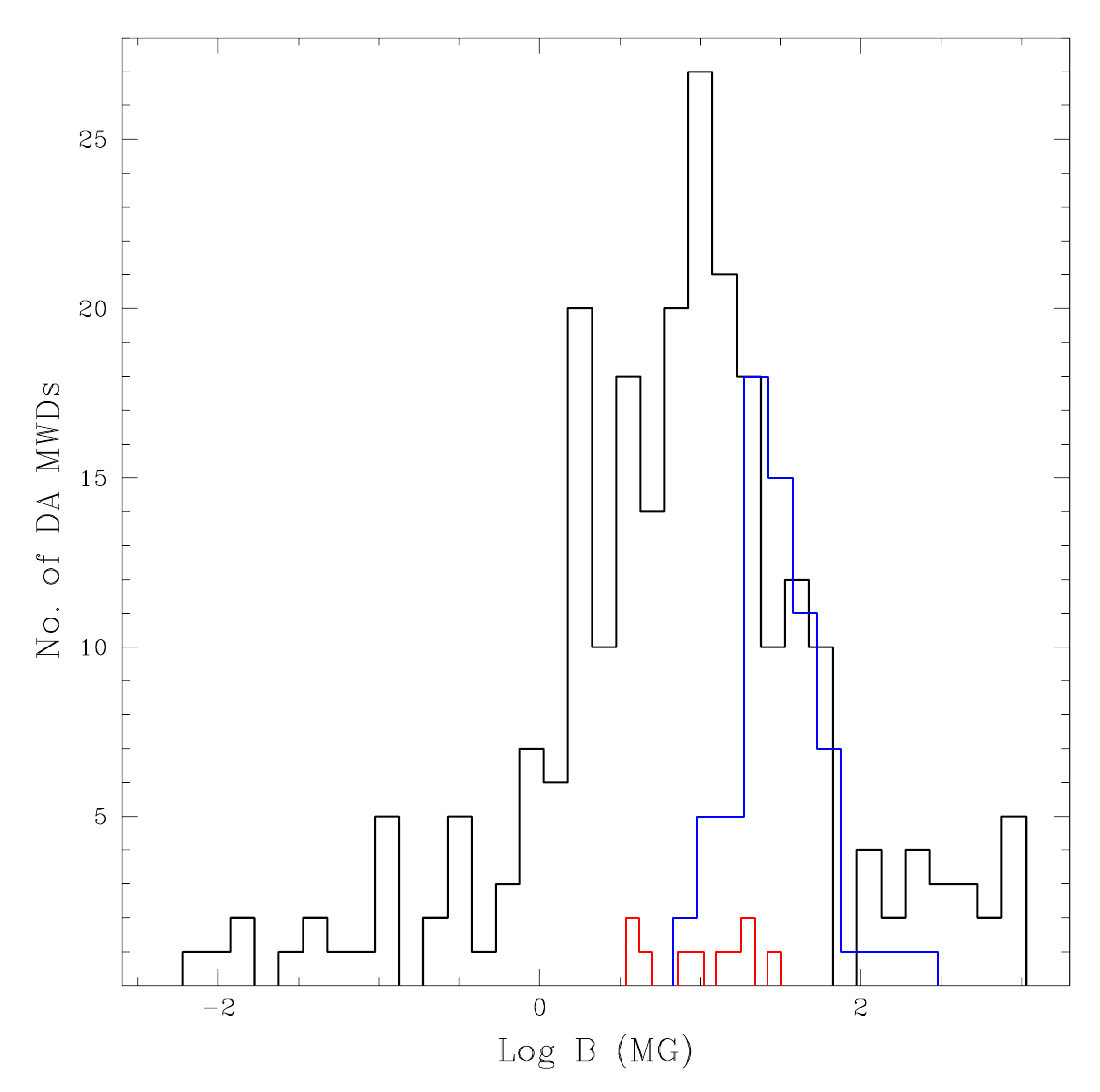}
\caption{Distribución de la intensidad del campo magnético en EBMs aisladas (negro), polares (azul), y polares intermedios (rojo) \cite{2015SSRv..191..111F}.}
\label{fig:Bdist}
\end{center}
\end{figure}

La intensidad del campo magnético superficial en las EBMs, varía entre \mbox{$10^3$ G} y \mbox{$10^9$ G}, y distribuye según la Figura \ref{fig:Bdist}. La topología de los mismos puede llegar a ser bastante complicada, aunque suele asumirse una configuración dipolar \cite[y referencias encontradas en él]{2015SSRv..191..111F}.

\begin{figure}[ht]
\begin{center}
	\includegraphics[width=.7\textwidth]{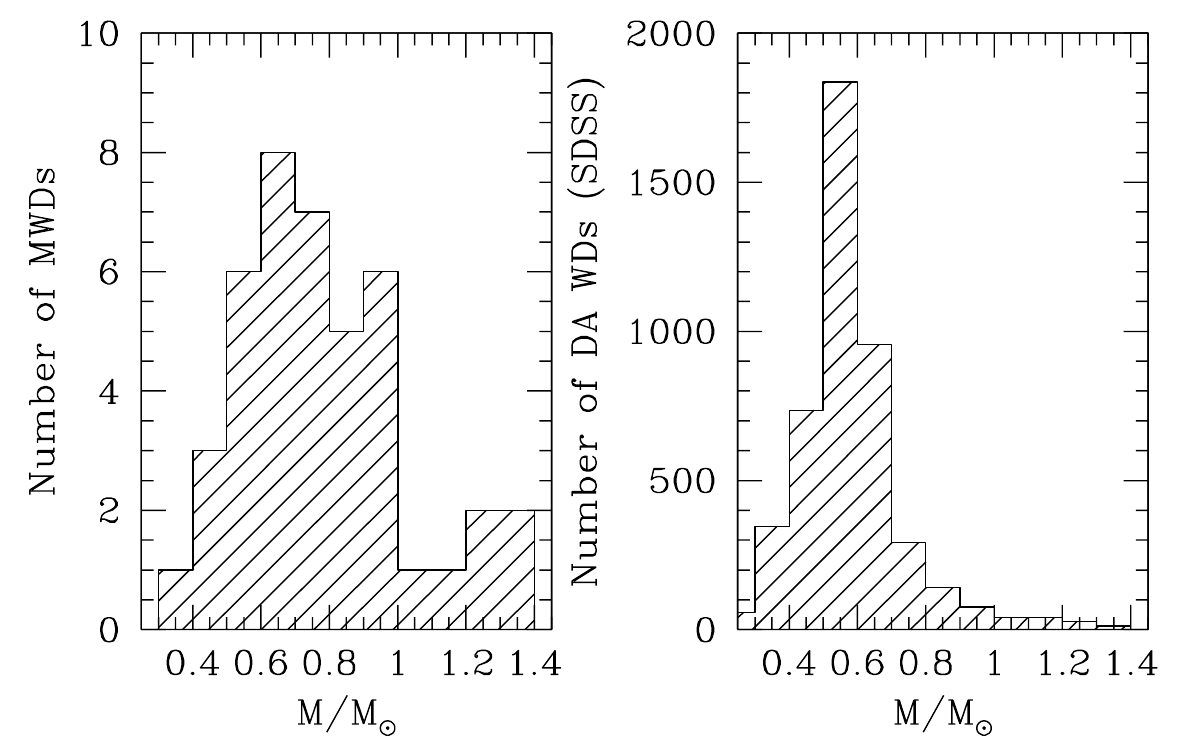}
\caption{Distribución de masa para EBMs \cite{2010AIPC..1273..378F}, y EBs no magnéticas \cite{2007MNRAS..375..1315K}.}
\label{fig:mdist}
\end{center}
\end{figure}

Un resultado bien establecido es que las EBMs  tienen masas mayores que las EBs que no presentan campos magnéticos, cuya masa promedio es $(0.663 \pm 0.136)\,M_\odot$. Para EBMs aisladas con campos mayores que $10^6$ G, la distribución de masa tiene una masa media de $(0.784\pm0.047)\,M_\odot$ y exhibe una cola que se extiende hasta el límite de Chandrasekhar (Figura \ref{fig:mdist}).


\chapter{Propiedades termodinámicas para un gas degenerado de electrones} \label{cap2}

Las propiedades del gas degenerado de electrones han sido extensamente estudiadas \cite{1983bhwd.book.....S,
camenzind2007compact}. Estas son extremadamente importantes cuando se estudia las enanas blancas, ya que es la presión de dicho gas lo que impide el colapso gravitacional de la estrella.

En este capítulo, partiendo del tensor de energía-momento, describimos las propiedades termodinámicas de un gas de electrones, tanto en ausencia como en presencia de campo magnético, obteniendo ecuaciones de estado apropiadas en cada caso. En particular, cuando el campo magnético es distinto de cero, se analiza la anisotropía que este produce en las ecuaciones de estado.

\section{Tensor de energía-momento}\label{c2TEM}

En aras de describir las propiedades termodinámicas del gas de electrones consideraremos inicialmente el tensor de energía-momento de un gas de fermiones cargados en presencia de un campo magnético, cuya obtención siguiendo los métodos tradicionales de la teoría cuántica de campos a temperatura finita \cite{berestetskii1981teoria,huang2008quantum,2000PhRvL..84.5261C}, presentamos de forma resumida.

Partiremos de la definición:
\begin{equation}
	\label{tem_def}
	\mathcal{T}_{\mu \nu} = \frac{\partial \mathcal{L}}{\partial A_{\mu,\nu}}  A_{\mu,\nu} - \delta_{\mu \nu} \mathcal{L},
\end{equation}
donde
	\begin{equation}
		\mathcal{L} = \overline{\psi} \left[\left(i\partial_\mu-eA_\mu\right)\gamma_\mu - m\right] \psi+ \frac{1}{4} F_{\mu\nu} F^{\mu\nu}
	\end{equation}
es la densidad Lagrangiana de  Dirac, $\gamma_\mu$ las matrices de Dirac, $\psi$ el campo fermiónico, $A_\mu$ el potencial del campo electromagnético, $F_{\mu\nu}$ el tensor dual de Maxwell, $m$ la masa del electrón y $e$ la carga del mismo.

El tensor de energía-momento macroscópico es el promedio estadístico $T_{\mu\nu} = \langle \mathcal{T}_{\mu \nu} \rangle$, donde se tiene en cuenta que el potencial termodinámico $\Omega$ pasa a ocupar formalmente el lugar de la densidad Lagrangiana:
\begin{equation}
	\label{omLag}
	\Omega = - \frac{1}{\beta} \ln \langle e^{\int_0^\beta d x_4 \int d^3 x \mathcal{L}(x_4, \vec x)} \rangle,
\end{equation}

\noindent siendo $\beta$ el inverso de la temperatura absoluta $T$. Los detalles de estos cálculos pueden verse en \cite{2000PhRvL..84.5261C}. 

Entonces, el tensor de energía-momento toma la forma  \cite{2000PhRvL..84.5261C}:
\begin{equation}\label{tensor_e-m}
T_{\mu\nu}=\left(T\frac{\partial{\Omega}}{\partial{T}}+\mu \frac{\partial{\Omega}}{
\partial{\mu}}\right)\delta_{4\mu}\,\delta_{\nu4}+4F_{\mu \alpha}F_{ \nu \alpha}\frac{\partial{\Omega}}{
\partial{F^{2}}}-\Omega\,\delta_{\mu\nu}.
\end{equation}
donde $\mu$ es el potencial químico de los electrones y positrones.

Como es conocido \cite{berestetskii1981teoria}, en un sistema de referencia comóvil con el gas, el tensor de energía–
momento tiene forma diagonal, y las presiones se obtienen de las componentes diagonales espaciales del mismo, $T_{11}$, $T_{22}$, y $T_{33}$, mientras que la densidad de energía proviene de la componente  $T_{44}$.
Si el sistema está en presencia de un campo magnético  $\vec B$ orientado en la dirección $x_3$ ($x_\mu = (x_4, \vec x)$) las componentes no nulas del tensor (\ref{tensor_e-m}) \cite{2000PhRvL..84.5261C,Felipe2005ChJAA} toman la forma:
\begin{subequations}
\begin{align}  \label{componentes de T_a}
T_{44}&=E=TS+\mu N+\Omega, \\
T_{11}&=T_{22}=P_{\perp}=-\Omega-B\mathcal{M},  \label{componentes de T_b}  \\
T_{33}&= P_{\|}=-\Omega.  \label{componentes de T_c}
\end{align}
\end{subequations}

Las magnitudes $E$, $N$  y $\mathcal{M}$ son respectivamente la densidad de energía, la densidad de partículas y la magnetización. Además, en estas expresiones aparece una presión en la dirección paralela $P_\parallel$ y otra en la perpendicular $P_\perp$ al campo magnético, debido a la ruptura de la simetría rotacional $O(3)$ que introduce $B$.

En el límite de campo magnético cero, se recupera la forma del tensor del fluido perfecto:
\begin{equation}
 T_{\mu\nu}=P\delta_{\mu\nu}-(P+E)\delta_{4\mu}\,\delta_{\nu4},
\end{equation}
cuyas  componentes no nulas espaciales se corresponden con la presión  y la $T_{44}$ es la energía:
\begin{subequations}
\begin{align}
T_{11}&=T_{22}= T_{33} = P=-\Omega,  \label{componentes de T_1}  \\
T_{44}&=-E=-TS-\mu N-\Omega,
 \label{componentes de T_4}
\end{align}
\end{subequations}

Es importante destacar que la contribución  de Maxwell $F_{\mu\nu} F^{\mu\nu}$,  también da lugar a presiones anisotrópicas $P_\perp=-P_{\parallel}= P_{M}=B^2/8\pi$ con densidad de energía  $\varepsilon=P_{M}$ \cite{Ferrer2010PhRvC}. No obstante, a densidades de energía típicas de las enanas blancas \mbox{$E \sim (10^{-7} - 10^{-3})$ MeV fm$^{-\textrm{3}}$}, los efectos de $P_{M}$ no serán notables para campos magnéticos $B<10^{14}$ G, por lo que no tendremos en cuenta este término \cite{2007PhR...442..109L}.

Para obtener las presiones y la energía, tenemos que calcular el potencial termodinámico del sistema. A continuación, determinamos el potencial en ausencia de campo magnético ($B=0$), con el que se obtienen las ecuaciones de estado correspondientes; y luego realizamos el mismo procedimiento para el campo magnético diferente de cero ($B\neq0$).

\section{Potencial termodinámico a campo magnético cero} \label{c2ptm}

Con el objetivo de obtener las propiedades termodinámicas del gas de electrones partiremos de la expresión general para el potencial termodinámico:
\begin{equation}\label{Grand-Potential-Def}
\Omega(\mu,T,0)=\frac{1}{\beta V} \text{Tr} \ln Z=\frac{1}{\beta V}\sum_{p_4}\int \frac{d^3p}{(2\pi)^3}\ln \det G^{-1}(\overline{p})
\end{equation}
siendo  $Z$ la función de partición del ensemble gran canónico y $G^{-1}(\overline{p})$ el propagador  fermiónico en el espacio de los momentos: 
\begin{equation}
G^{-1}(\overline{p})=[{\overline{p}}\cdot\gamma-m],
\label{Inv-Propagator-Def}
\end{equation}
donde $\overline{p}=(i p_4-\mu,p_1,p_2,p_3)$.
La traza y el logaritmo en (\ref{Grand-Potential-Def}) se toman en el sentido funcional, quedando entonces:

\begin{equation}\label{Thermo-Potential_B0}
\Omega (\mu,T,0)=-\frac{1}{\beta}\int_{0}^\infty \frac{d^3p}{(2\pi)^3}\sum_{p_4}\ln \left[(p_4+i\mu)^2+\epsilon^2\right],
\end{equation}
con el espectro:
\begin{equation}\label{Disp-Rel_B0}
\epsilon=\sqrt{p^2+m^2}.
\end{equation}

Partiendo de la ecuación (\ref{Thermo-Potential_B0}), realizamos la suma por $p_4$ siguiendo el procedimiento de Matsubara \cite{fradkin67}:
 \begin{equation}\label{Matsubara}
	 p_4=\frac{(2n+1)\pi}{\beta}, \quad n=0,\pm1,\pm2,...
\end{equation}
y obtenemos el potencial termodinámico del sistema:
\begin{equation}\label{omega_n}
\Omega(\mu,T,0)=-\frac{1}{4\pi^2\beta}\int d^3 p \; \ln \left[\left(1+e^{\beta (\epsilon+\mu)}\right)\left(1+e^{\beta(\epsilon-\mu)}\right)\right],
\end{equation}
donde  $[(1+e^{(\epsilon\mp\mu)\beta})]^{-1}$
son las distribuciones asociadas a las partículas y antipartículas  respectivamente.

Si tenemos en cuenta que las temperaturas típicas de las enanas blancas ($10^7$ K $\sim 10^{-3}$ MeV),
 son mucho menores que la temperatura de Fermi  $T_F= 10^9$ K  ($T/T_F =10^{-2}\ll1$), se justifica tomar el límite degenerado $T=0$ ($\beta \rightarrow \infty$). En este caso, el potencial termodinámico toma la forma:
\begin{equation}
	\label{TH_Pot_0}
	\Omega (\mu, 0, 0)= - \frac{1}{4\pi^3} \int d^3 \vec p \left(\mu - \sqrt{p^2+m^2}\right)\Theta \left(\mu - \sqrt{p^2+m^2}\right),
\end{equation}
donde la contribución de los positrones se hace cero y la función de distribución de los electrones se transforma en la función paso unitario $\Theta \left(\mu - \epsilon\right)$:
\begin{equation*}
	\Theta \left(\mu - \epsilon\right)= \begin{cases}
				1, & \mu>\epsilon;\\
				0, & \mu<\epsilon.
			\end{cases}
\end{equation*}

 Al resolver la integral planteada en (\ref{TH_Pot_0}) como se explica en el Apéndice \ref{appA}, resulta:
 \begin{equation}
	\label{TH_Pot_0:2}
	\Omega (\mu, 0, 0) = - \frac{m^4}{4\pi^2} \left[\frac{\mu \sqrt{\mu^2-m^2}}{3 m^2}\left(\frac{\mu^2}{m^2}-\frac{5}{2}\right) + \frac{1}{2} \ln\left(\frac{\mu+\sqrt{\mu^2-m^2}}{m}\right)\right].
\end{equation}

\section{Ecuaciones de estado para campo magnético cero}\label{c2s2}

A partir de la expresión (\ref{TH_Pot_0:2}) para el potencial termodinámico, podemos hallar todas las magnitudes termodinámicas del sistema. En particular la densidad de partículas:
 \begin{align}
 	\label{TH_Pot_0:N}
	 N (\mu, 0, 0) &= -\frac{d \Omega (\mu, 0, 0)}{d\mu}, \\  	\label{TH_Pot_0:N1}
	 N(x,0,0) &= \frac{m^3}{3\pi^2} x^3,
\end{align}
donde hemos introducido el momentum de Fermi adimensional $x=p_F/m$, con $p_F=\sqrt{\mu^2 -m^2}$.

Luego, obtenemos la energía y la presión del sistema:
\begin{subequations} \label{TH_Pot_0:EoS}
\begin{align}	\label{eq:e0}
	E (\mu, 0, 0) &= \Omega (\mu, 0, 0) + \mu \, N (\mu, 0, 0),\\ \label{eq:P0}
	P (\mu, 0, 0) &= -\Omega (\mu, 0, 0),
\end{align}
\end{subequations}
cuyas expresiones determinan la ecuación de estado.

En función del momento de Fermi, la ecuación (\ref{eq:e0}) se transforma en:
\begin{subequations}
\begin{align}	\label{TH_Pot_0:EoS_E}
	E (x, 0, 0)&= m^4 \chi(x), \\
	\chi(x)&=\frac{1}{8\pi^2} \left[x\sqrt{x^2+1}\left(2x^2+1\right) -\ln \left(\sqrt{x^2+1}+x\right)\right],
\end{align}
\end{subequations}
	
y (\ref{eq:P0}) en:
\begin{subequations}  \label{TH_Pot_0:EoS_P}
\begin{align}	
	P (x, 0, 0) &= m^4\Phi(x),\\
	\Phi (x) &=\frac{1}{8\pi^2} \left[x\sqrt{x^2+1}\left(\frac{2}{3}x^2-1\right) +\ln \left(\sqrt{x^2+1}+x\right)\right].
\end{align}
\end{subequations}

Para \textbf{electrones relativistas ($x\gg1$)}, se puede desarrollar las funciones $\chi (x)$ y $\Phi(x)$  en serie de potencias de $x$ en torno a infinito, quedando:
\begin{subequations}
\begin{align}	 \label{TH_Pot_0:staylorr}
	\chi(x)&=\frac{1}{4\pi^2}\left[x^4 +x^2-\frac{1}{2}\ln \left(2 x\right)+\frac{1}{8}+\ldots\right], \\
	\Phi (x) &=\frac{1}{12\pi^2}\left[x^4 -x^2+\frac{3}{2}\ln \left(2 x\right)-\frac{7}{8}+\ldots\right].
\end{align}
\end{subequations}

Por otra parte,
para los \textbf{electrones no relativistas ($x\ll1$)}, es posible proceder expresando $\chi (x)$ y $\Phi(x)$  como una serie de potencias de $x$ en torno a cero:
\begin{subequations}
\begin{align} 	 \label{TH_Pot_0:taylornr}
	\chi(x)&=\frac{1}{3\pi^2}\left[x^3 +\frac{3}{10}x^5-\frac{3}{56}x^7+\ldots\right],\\
	\Phi (x) &= \frac{1}{15\pi^2}\left[x^5 -\frac{5}{14}x^7+\frac{5}{24}x^9+\ldots\right].
\end{align}
\end{subequations}

\section{Potencial termodinámico en presencia de campo magnético}

Consideremos un gas de electrones magnetizado a temperatura finita.  En presencia de un campo magnético constante $\vec B$ dirigido según el eje $x_3$.  Podemos decir que  el gas de electrones magnetizado experimenta tres cambios importantes:

\begin{enumerate}
\item El propagador fermiónico se modifica siendo $G^{-1}_l(\overline{p})$, con  $\overline{p}=(i p_4-\mu,0,\sqrt{2eBl},p_3)$.
Consecuentemente la relación de dispersión de los electrones \cite{1949PhRv...76..828J, berestetskii1981teoria} que resulta de resolver la ecuación de Dirac:
\begin{equation}
	\label{TH_Pot_B:spectra}
	\epsilon_l = \sqrt{p^2_3+m^2 + 2e B l},
\end{equation}

muestra cuantización de Landau de las frecuencias ciclotrónicas. Hemos designado con $l$ los niveles de Landau.

\item La densidad de estados se vuelve proporcional al campo magnético de manera que la integral por los momenta en (\ref{omega_n}) se transforma según:
\begin{equation}
	\label{TH_Pot_B:trans}
	2 \int \frac{d^3\vec p}{(2\pi)^3} \longrightarrow 2 \sum_{l=0}^\infty g(l) \int \frac{eB}{(2\pi)^2} dp_3,
\end{equation}
donde $g(l)=(2-\delta_{l0})$ considera la doble degeneración del espín para todos los niveles de Landau excepto el $l=0$.

\item Como ya vimos en la Sección \ref{c2TEM}, se rompe la  simetría rotacional  $O(3)$, dando lugar a anisotropías en el tensor de energía-momentum. Esto produce una separación de la presión  en dos componentes distinguibles: una a lo largo del campo (la presión paralela) y otra en la dirección transversal (la presión perpendicular). Como resultado, las ecuaciones de estado son anisotrópicas.
\end{enumerate}

Por consiguiente, el potencial termodinámico toma la forma:
\begin{equation}
	\label{TH_Pot_B:0}
	\Omega (\mu, T, B)=  - \frac{e B}{2\pi^2 } \int_{-\infty}^\infty  dp_3 \sum_{l=0}^\infty g(l) \left\{\epsilon_l+  \frac{1}{\beta} \ln  \left[\left(1+e^{\beta (\epsilon_l+\mu)}\right)\left(1+e^{\beta(\epsilon_l-\mu)}\right)\right]\right\}.
\end{equation}

En la ecuación (\ref{TH_Pot_B:0}), se pueden identificar dos términos: la contribución del vacío (\ref{TH_Pot_B:vac}), y la parte originada por el efecto estadístico del gas de electrones (\ref{TH_Pot_B:st}) que depende de $\mu$, $T$ y $B$.

\begin{subequations}
\begin{align}
	\label{TH_Pot_B:vac}
	\Omega_{\text{vac}} (0, 0, B) &=  - \frac{e B}{2\pi^2 } \int_{-\infty}^\infty  dp_3 \sum_{l=0}^\infty g(l) \,\epsilon_l, \\
	\label{TH_Pot_B:st}
	\Omega_{\text{est}} (\mu, T, B) &=  - \frac{e B}{4\pi^2 \beta} \int_{-\infty}^\infty  dp_3 \sum_{l=0}^\infty g(l) \ln \left[\left(1+e^{\beta (\epsilon_l+\mu)}\right)\left(1+e^{\beta(\epsilon_l-\mu)}\right)\right].
	\end{align}
\end{subequations}

Por tanto,
\begin{equation}
	\label{TH_Pot_B:0t}
	\Omega (\mu, T, B)=  \Omega_{\text{vac}} (0, 0, B) +\Omega_{\text{est}} (\mu, T, B).
\end{equation}

Como argumentamos en la Sección (\ref{c2ptm}), nuestro interés es modelar enanas blancas magnetizadas. Para ellas, la aproximación del gas degenerado es suficiente, de manera que el término estadístico del potencial (\ref{TH_Pot_B:0t}) queda:
\begin{equation}
	\label{TH_Pot_B:deg}
	\Omega_{\text{est}} (\mu, 0, B) = - \frac{eB}{4\pi^2 } \sum_{l=0}^\infty (2-\delta_{l0}) \int_{-\infty}^\infty dp_3\,
	(\mu -\epsilon_l)\Theta(\mu -\epsilon_l)
\end{equation}

\subsection{Límite de campo magnético débil ($B<B_c$)}

Se define el campo magnético crítico de Schwinger $B_c = m^2/e = 4.41\times10^{13}$ G, como el campo magnético a partir del cual la energía ciclotrónica de los electrones $eB/m^2$ es comparable con su masa en reposo.
Este valor permite determinar dos regímenes: el llamado límite de campo magnético débil ($B<B_c$), y el límite de campo magnético fuerte $B>B_c$.

Recordando que los campos magnéticos superficiales de las EBs están entre $10^3$ G y $10^9$ G, y que el valor del campo magnético en el centro de estas estrellas según el teorema del Virial escalar puede llegar hasta  $10^{12}$ G, en nuestro estudio tomaremos el límite de campo magnético débil.

\subsubsection{Contribución estadística}

En la aproximación de campo magnético débil $(B<B_c)$, la separación entre los niveles de Landau \mbox{$(\sim eB)$} es pequeña. Por tanto, podemos considerar el espectro discreto (\ref{TH_Pot_B:spectra}) como un continuo, y reemplazar la suma en $l$ por una integral mediante la fórmula de Euler-MacLaurin \cite{1972hmfw.book.....A}:
\begin{multline}
	\label{EMcL}
	eB \sum_{l=0}^\infty (2-\delta_{l0}) f(2eBl) \approx \int_0^\infty (2eB) f(2eBl)dl + (eB) f(\infty)\\
	+ \sum_{k=1}^\infty \frac{(2eB)^{2k} }{(2k)!} B_{2k} \left[f^{2k-1}(\infty)-f^{2k-1}(0)\right],
\end{multline}
donde el potencial se expresa como función de las potencias del campo magnético, y hemos definido:
\begin{equation}
	\label{TH_Pot_B:f}
	f(2eBl)=\left(\mu - \sqrt{p_3^2+m^2+2eBl}\right)\Theta \left(\mu - \sqrt{p_3^2+m^2+2eBl}\right).
\end{equation}

La validez de esta aproximación se ilustra en la Tabla \ref{tab:lmax}. En ella calculamos el número máximo de niveles de Landau como función del campo magnético y el potencial químico:
\begin{equation}
	\label{eq:lmax}
	l_{max} = \textrm{I}\left[ \frac{\mu^2-m^2}{e B}\right].
\end{equation}

Se denota  como $ \textrm{I}\left[x\right]$ la función parte entera de $x$. Este nivel máximo se obtiene al tener en cuenta que el momentum de Fermi debe ser positivo en la expresión del potencial $\Omega_{\text{est}} (\mu, 0, B)$  \cite{Felipe2005ChJAA}.

\begin{table}[h]
\begin{center}
\begin{tabular}{ccccc} \toprule
 			& \multicolumn{4}{c}{$l_{max} $} \\ \cmidrule{2-5}
 	$\mu$ (MeV)	&  $B=5\times10^{10}$ G & $B=5\times10^{11}$ G & $B=5\times10^{12}$ G &$B=10^{13}$ G \\ \midrule
$ 0.722663$ & 2 200 & 219 & 22 & 2 \\
 $1.61592$ & 19 800 & 1 979 & 198 & 19 \\
$ 2.6056$ & 55 000 & 5 499 & 550 & 55 \\
 $3.61332$ & 107 800 & 10 779 & 1 078 & 107 \\
 $4.6273$ & 178 200 & 17 819 & 1 782 & 178 \\
 $5.64418$ & 266 200 & 26 619 & 2 662 & 266 \\
 $6.66262$ & 371 800 & 37 179 & 3 718 & 371 \\
 $7.68201$ & 495 000 & 49 499 & 4 950 & 495 \\
 $8.70202$ & 635 800 & 63 579 & 6 358 & 635 \\
 $9.72244$ & 794 200 & 79 419 & 7 942 & 794 \\
 $10.2328$ & 880 000 & 87 999 & 8 800 & 880 \\ \bottomrule
\end{tabular}
\end{center}
\caption{Nivel de Landau máximo para valores de $\mu$ y $B$ dados. En la aproximación de campo magnético débil ($B<B_c$), hay suficientes niveles como para justificar el uso de la fórmula de Euler-MacLaurin para transformar la suma por los niveles de Landau en una integral.}
\label{tab:lmax}
\end{table}

Procediendo como se explica en el Apéndice \ref{appA}, obtenemos la parte estadística del potencial termodinámico del gas degenerado de electrones en presencia de campo magnético:
\begin{equation}
	\label{TH_Pot_B:st4}
	\Omega_{\text{est}}(\mu, 0, B)   = \Omega(\mu, 0, 0)   + \Omega_B,
\end{equation}
 donde el término asociado al campo magnético tiene la forma:
\begin{equation}
	\label{TH_Pot_B:st3}
	\Omega_B  = - \frac{m^4}{12\pi^2}\left[\frac{B}{B_c}\right]^2\ln\left(\frac{\mu+\sqrt{\mu^2-m^2}}{m}\right).
\end{equation}

\subsubsection{Contribución del vacío}

Retomemos el potencial  (\ref{TH_Pot_B:0}). El término de vacío (\ref{TH_Pot_B:vac}) presenta una divergencia ultravioleta no dependiente del campo magnético que se evita mediante una renormalización (ver el procedimiento detallado en \cite{berestetskii1981teoria}).

Después de renormalizar, queda la bien conocida expresión de Schwinger \cite{PhysRev.82.664}:
\begin{equation}\label{Thermo-Potential-5}
\Omega_{\text{vac}} (0, 0, B)=-\frac{1}{8\pi^2}\int_{0}^\infty \frac{ds}{s^3}exp(-m^2s)\left(esB\coth (esB)-1-\frac{1}{3}(esB)^2 \right),
\end{equation}
que en el límite de campo magnético débil es calculada en \cite{PhysRevD.91.085041} y se reduce a:
\begin{equation} \label{TH_Pot_B:vac1}
	\Omega_{\text{vac}} (0, 0, B) = \frac{m^4}{90(2\pi)^2} \left(\frac{B}{B_c}\right)^4.
\end{equation}

Teniendo en cuenta que estamos considerando la aproximación de campo magnético débil \mbox{($B\!<\!B_c$)}, podemos notar que $\Omega_{\text{vac}} (0, 0, B) \ll \Omega_B$. Consecuentemente, el término dominante en este régimen de campo magnético es el estadístico, y el de vacío será despreciado en nuestros cálculos.

\section{Ecuaciones de estado en el régimen de campo magnético débil}

Considerando el potencial termodinámico (\ref{TH_Pot_B:st4}) del sistema, las ecuaciones de estado quedan:
\begin{subequations} \label{TH_Pot_B:EoS}
\begin{align}
	E(\mu, 0, B)   &= \Omega_{\text{est}}(\mu, 0, B)   + \mu \, N(\mu, 0, B) ,  \\
	P_\parallel (\mu, 0, B)  &= - \Omega_{\text{est}}(\mu, 0, B) ,  \\
	P_\perp (\mu, 0, B)  &= - \Omega_{\text{est}}(\mu, 0, B) - B\mathcal{M}(\mu, 0, B) .
\end{align}
\end{subequations}
donde la densidad de partículas  y la magnetización se determinan según:
\begin{eqnarray}
	 N(\mu, 0, B)  &=& - \frac{\partial \Omega_{\text{est}}(\mu, 0, B)  }{\partial \mu}, \\
	\mathcal{M} (\mu, 0, B) & = &- \frac{\partial \Omega_{\text{est}}(\mu, 0, B)}{\partial B},
\end{eqnarray}

En función del momento de Fermi adimensional, tenemos:
\begin{eqnarray} \label{TH_Pot_B:N}
	 N(x, 0, B)  &= & N(x,0,0) + N_B, \\
	 N_B &=& \frac{m^3}{12\pi^2} \left[\frac{B}{B_c}\right]^2\frac{1}{x}, \\
	  \label{TH_Pot_B:M}
	\mathcal{M}(x, 0, B) &=& \frac{m^4}{6\pi^2} \frac{B}{B_c^2} \ln \left(x+\sqrt{x^2+1}\right).
 \end{eqnarray}

Luego, sustituyendo la forma explícita del potencial termodinámico $\Omega_{\text{est}}$, la densidad de partículas $N$ y la magnetización $\mathcal{M}$ dadas por las expresiones (\ref{TH_Pot_B:st4}), (\ref{TH_Pot_B:N}) y (\ref{TH_Pot_B:M}) respectivamente, podemos escribir las ecuaciones de estado, en función del momento de Fermi adimensional:
\begin{subequations} \label{TH_Pot_B:EoS2}
\begin{align}\label{TH_Pot_B:EoS2i}
	E(x, 0, B) &=  E(x, 0, 0) + \left(\Omega_B + m\, N_B\sqrt{x^2+1} \right), \\
	P_\parallel (x, 0, B) &= P (x, 0,0)  - \Omega_B, \\
	P_\perp (x, 0, B) &= P (x, 0, 0)  + \Omega_B.
\end{align}
\end{subequations}

Nótese que las expresiones anteriores muestran la energía y presiones como la suma de un término no dependiente del campo magnético y otro que sí es función de este.


\chapter{Ecuaciones de estado para enanas blancas}
\label{cap3}

Las ecuaciones de estado constituyen una parte fundamental en los estudios astrofísicos sobre objetos compactos. Describen la microfísica del sistema a partir de la relación entre la presión y la densidad de masa o la densidad de energía del sistema, lo cual indica cuánto resiste la materia en cuestión ante una compresión. En este sentido, los modelos para caracterizar la estructura de dichos objetos compactos se nutren de ellas para completar la información que aportan a las ecuaciones de equilibrio hidrodinámico.

Este capítulo abarca la parte original de la tesis, pues introduce la corrección electrostática debida a la interacción entre las partículas, electrones e iones, que componen la enana blanca, y la no uniformidad en la distribución de los nucleones. A modo introductorio, presentamos las ecuaciones de estado para un gas no interactuante.

\section{Ecuaciones de estado en condiciones de equilibrio estelar}

Las condiciones de equilibrio para las EBs establecen la neutralidad de carga (\ref{EoS_nc}) y la conservación del número de bariones (\ref{EoS_cb}):
\begin{eqnarray} \label{EoS_nc}
	N_e &=& N_p\\ \label{EoS_cb}
	N_b &=& N_n+N_p
\end{eqnarray}

Debido a que los protones y neutrones son aproximadamente $10^3$ veces más masivos que los electrones, se comportan como fermiones no relativistas. Por tanto, en la expresión para la densidad de energía (masa) debe considerarse la masa en reposo del nucleón $m_N = 931.494\, \textrm{MeV}$, mientras que el aporte del mismo a la presión en el caso degenerado es despreciable. Esto permite afirmar que es precisamente la presión del gas degenerado de electrones lo que compensa la atracción gravitacional, evitando el colapso de la estrella.

\subsection{Enanas blancas}

Al considerar la masa en reposo de los iones y el gas libre de electrones, la densidad de energía para enanas blancas en ausencia de campo magnético es:
\begin{equation} 
\label{EoS_0:E}
	E (\mu,0,0) = \Omega(\mu,0,0) + \mu\, N(\mu,0,0) +  \frac{A}{Z} m_N N(\mu,0,0),
\end{equation}
donde $Z/A$ es el número de electrones por barión.

Esto nos permite definir la función:
\begin{equation}
	\chi \left( x,A/Z\right) = \chi (x) + \frac{1}{3\pi^2}\frac{A}{Z}\frac{m_N}{m} x^3,
\end{equation}
de manera que la densidad de energía queda:
\begin{equation} 
\label{EoS_0:E1}
	E_{A/Z} (x,0,0) = m^4 \chi \left( x,A/Z\right).
\end{equation}

\subsubsection{Ecuación politrópica}

Las ecuaciones de estado formada por la energía (\ref{EoS_0:E1}) y  la presión (\ref{TH_Pot_0:EoS_P}) son paramétricas en el momentum de Fermi adimensional. Para las enanas blancas, usualmente se considera la condición $m_N\gg m$, y se escribe una ecuación de estado politrópica:
\begin{equation} \label{eq:polit}
	P=\kappa E^\Gamma,
\end{equation}
con 
\begin{equation} \label{eq:politk}
	\kappa(A/Z) = \frac{1}{3 \Gamma} \left(\frac{3\pi^2}{m^4}\right)^{\Gamma-1}\!\!\left(\frac{A}{Z} \frac{m}{m_N}\right)^{\Gamma}\!.
\end{equation}

Para \textbf{electrones no relativistas ($x\ll1$)},  los parámetros $\Gamma$ y $\kappa(A/Z)$ tienen los valores siguientes:
\begin{subequations} 
\begin{align}
	\Gamma &= \frac{5}{3},\\
	\kappa(A/Z) &=\frac{1}{5} \left(\frac{3\pi^2}{m^4}\right)^{2/3}\!\!\left(\frac{A}{Z}\frac{m}{m_N}\right)^{5/3}\!= \frac{4.2162\times 10^{-5}}{\textrm{MeV}^{8/3}}\left(\frac{A}{Z} \right)^{5/3}.
\end{align}
\end{subequations}

En el caso de \textbf{electrones relativistas ($x\gg1$)}, los valores de dichos parámetros serían:
\begin{subequations} 
\begin{align}
	\Gamma &= \frac{4}{3}, \\
	\kappa(A/Z) &= \frac{1}{4} \left(\frac{3\pi^2}{m^4}\right)^{1/3}\!\!\left(\frac{A}{Z} \frac{m}{m_N}\right)^{4/3}\!= \frac{8.5017\times 10^{-5}}{\textrm{MeV}^{4/3}}\left(\frac{A}{Z}\right)^{4/3}.
\end{align}
\end{subequations}

A pesar de que no la utilizaremos en el presente trabajo, esta aproximación es muy usada, pues facilita los cálculos tanto para las ecuaciones de estructura Newtonianas como para las obtenidas usando la TGR.  Específicamente en el caso Newtoniano, el empleo de la ecuación de estado politrópica conduce a la conocida ecuación de Lane-Emden \cite{1983bhwd.book.....S}.

\subsection{Enanas blancas magnetizadas en el régimen de campo débil ($B<B_c$)}

En la aproximación de campo débil, si se tiene en cuenta la energía en reposo de los iones, la energía (\ref{TH_Pot_B:EoS2i}) del gas libre magnetizado de electrones toma la forma:
\begin{equation} 
	E (x,0,B)= E_{A/Z}(x,0,0) +\Omega_B+\left(m\sqrt{x^2+1} + \frac{A}{Z} m_N\right)N_B.
\end{equation}

Luego, podemos definir:
\begin{subequations} \label{eq:Eb}
\begin{align}
	E_{A/Z}(x,0,B) &= m^4 \,\chi (x,A/Z,B), \\ \label{eq:xib}
	\chi (x,A/Z,B) &=\chi (x,A/Z)+\frac{1}{12\pi^2} \left(\frac{B}{B_c}\right)^2\!\left[\sqrt{x^2+1} +\frac{m_N}{m x}\frac{A}{Z}-\ln\!\left(x+\sqrt{x^2+1}\right)\right].
\end{align}
\end{subequations}

Asimismo, podemos escribir las presiones como:
\begin{subequations} 	\label{eq:Pb}
\begin{align}
	P_{\parallel} (x,0,B)&= m^4 \, \Phi_{+} (x,B),\\ \label{eq:Pperpb}
	P_{\perp} (x,0,B)&= m^4 \,\Phi_{-} (x,B), \\ \label{eq:phib}
	\Phi_{\pm} (x,B) &= \frac{1}{8\pi^2} \left[\Phi (x) \pm \frac{2}{3}\left( \frac{B}{B_c}\right)^2 \ln \left(x+\sqrt{x^2+1}\right)\right].
\end{align}
\end{subequations}

\begin{figure}[h]
   \centering
   \includegraphics[width=.8\textwidth]{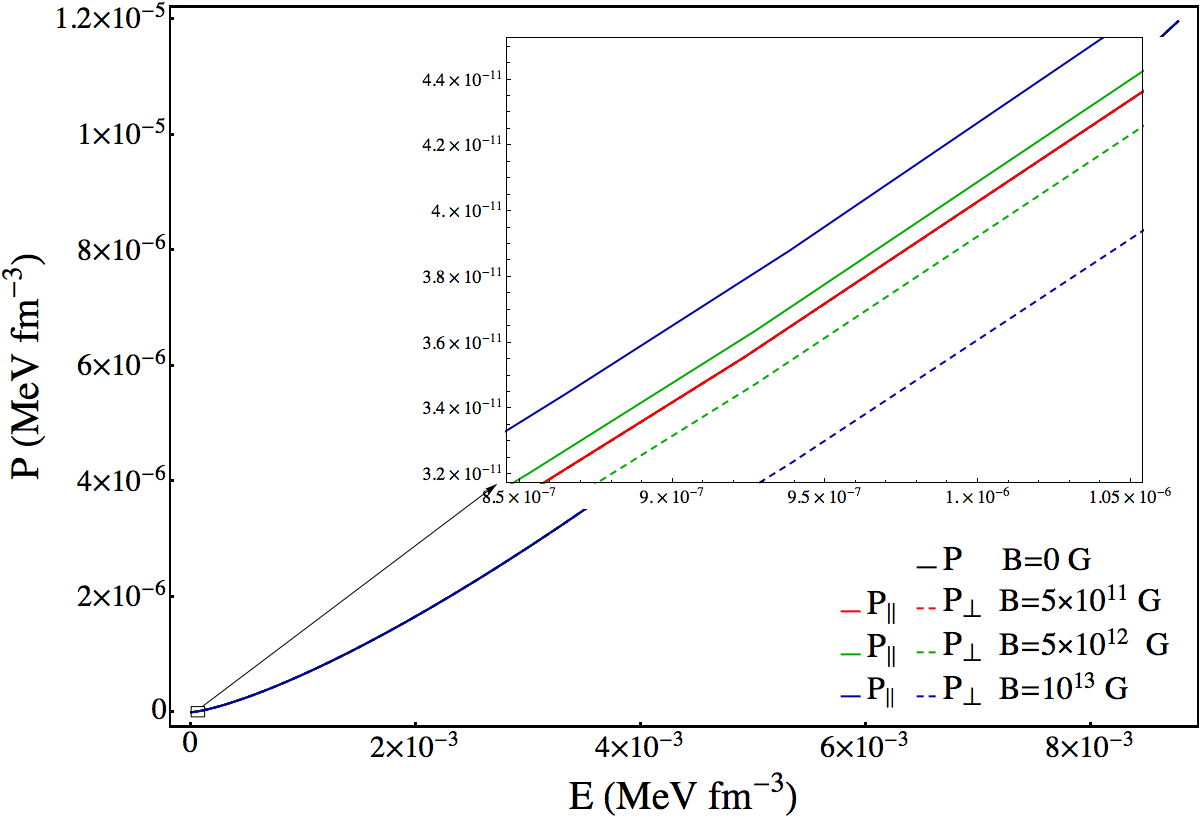} 
   \caption{Ecuaciones de estado para configuraciones con $A/Z=2$.}
   \label{fig:eosg}
\end{figure}

Las ecuaciones de estado resultantes, en función del momento de Fermi adimensional $x$, para un campo magnético dado, están determinadas por las ecuaciones (\ref{eq:Eb}) y (\ref{eq:Pb}), y se han graficado en la Figura \ref{fig:eosg}.

Para una mejor visualización del comportamiento de las ecuaciones obtenidas, desarrollamos  en serie de Taylor las funciones $\chi (x,A/Z,B)$ y $\Phi_{\pm} (x,B)$. En el caso de \textbf{electrones no relativistas ($x\ll1$)}, escribimos $\chi (x,A/Z,B)$ y $\Phi_{\pm} (x,B)$ como una serie de potencias de $x$ en torno a cero, donde:
\begin{subequations} 
\begin{align}
	\chi (x,A/Z,B) &=\chi (x,A/Z) + \frac{1}{12\pi^2}\left( \frac{B}{B_c}\right)^2\left[1 +\frac{m_N}{m x}\frac{A}{Z}+\frac{x}{2}-\frac{x^3}{24}+\ldots\right], \\
	\Phi_{\pm} (x,B) &= \Phi (x)\pm \frac{1}{15\pi^2}\frac{1}{4}\left( \frac{B}{B_c}\right)^2\left[5x -\frac{5x^3}{6}+\frac{3x^5}{8}\right]+\ldots
\end{align}
\end{subequations}

En cambio, para \textbf{electrones relativistas ($x\gg1$)}, el desarrollo en potencias de $x$ es respecto a infinito:
\begin{subequations} 
\begin{align}
	\chi (x,A/Z,B) &=\chi (x,A/Z) + \frac{1}{12\pi^2}\left( \frac{B}{B_c}\right)^2\left[- \frac{m_N}{m x}\frac{A}{Z}+\frac{x}{2}-\frac{x^3}{24}+\ldots\right], \\
	\Phi_{\pm} (x,B) &= \Phi (x)\pm\left( \frac{B}{B_c}\right)^2 \frac{\ln\left(2x\right) }{12\pi^2}+\ldots
\end{align}
\end{subequations}

\section{Ecuaciones de estado para enanas blancas magnéticas con partículas interactuantes}

Al obtener las ecuaciones de estado anteriores, se consideró que las partículas componentes de la enana blanca no interactúan entre sí, lo cual debe incluirse en un análisis más realista. En este sentido, existen varias correcciones, entre las que las electrostáticas y el decaimiento $\beta$ inverso son las más importantes.

La principal corrección electrostática tiene en cuenta que las cargas positivas no están distribuidas uniformemente en el gas, sino que se concentran en los núcleos ionizados.  Consecuentemente, la energía y la presión de los electrones disminuye, ya que la distancia media entre los electrones es mayor que la distancia media entre los electrones y los núcleos; y las fuerzas repulsivas son menores que las atractivas.

En un gas no degenerado, estos efectos Coulombianos se hacen más importantes a medida que aumenta la densidad de electrones, mientras que en el límite degenerado, los iones se localizan en una red cristalina que maximiza la distancia ión-ión. En la literatura se pueden encontrar diferentes aproximaciones para modelar la interacción Coulombiana, cada una de ellas da una ecuación de estado diferente \cite{shapiro2008black}. En este trabajo,  utilizaremos la corrección descrita en los artículos \cite{1982ApJ...253..839J,PhysRevD.90.043002,PhysRevD.92.023008}, que se basa en la ecuación de estado de Baym-Petthick-Sutherland (BPS) \cite{1971ApJ...170..299B}. 

\begin{figure}[h]
\centering
\includegraphics[width=.9\textwidth]{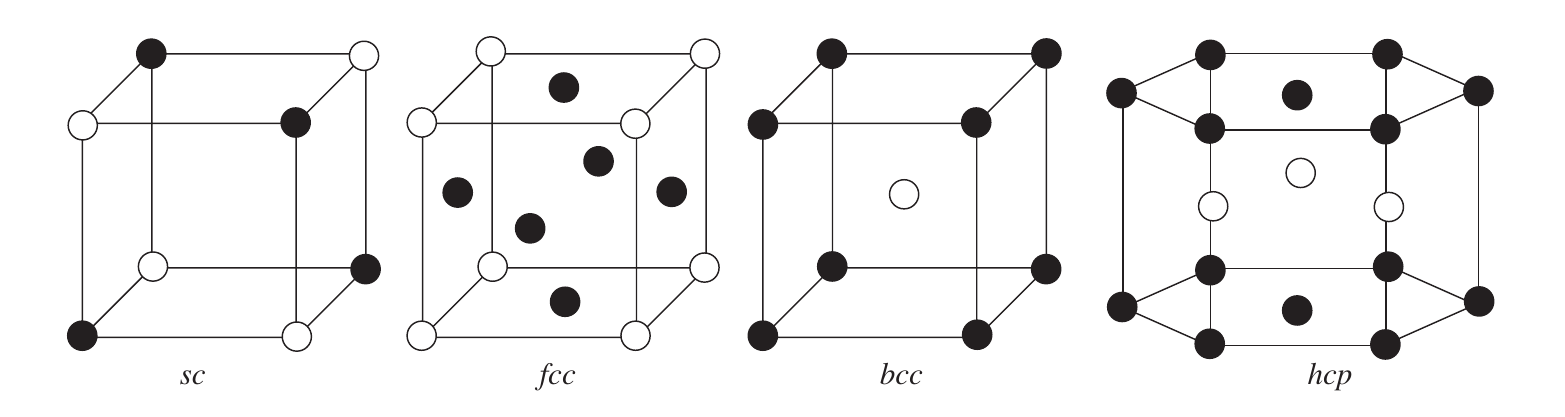}
\caption{Estructuras cristalinas heterogéneas cúbica simple (sc), cúbica centrada en las caras (fcc), cúbica centrada en el cuerpo (bcc) y hexagonal compacta (hcp). En todos los casos, los iones \mychemistryg{X}{A}{Z} se representan por círculos negros y los iones \mychemistryg{X$’$}{A$’$}{Z$’$} con círculos blancos.}
\label{fig:redt}
\end{figure}

Consideremos el sistema como un gas degenerado de electrones que rodea a una red de partículas puntuales compuesta por dos tipos de iones: \mychemistryg{X}{A}{Z} y \mychemistryg{X$’$}{A$’$}{Z$’$} (\Fref{fig:redt}).

En este caso, la densidad de energía de la red se escribe como:
\begin{equation} \label{eq:latticeE}
	\varepsilon_L = Ce^2 [N(x,0,B)]^{4/3} \,\frac{\eta Z^2 +\zeta Z’^2 +(1-\eta-\zeta)ZZ’}{(\xi Z + (1-\xi) Z’)^{4/3}},
\end{equation}
donde $\xi$ es la fracción de iones del tipo \mychemistryg{X}{A}{Z} y $\xi’=1-\xi$ la fracción de iones \mychemistryg{X$’$}{A$’$}{Z$’$}; $C$, $\zeta$ y $\eta$ son constantes que dependen del tipo de red escogida (Tabla \ref{tab:lcte}); y $N(x,0,B)$ es la densidad de electrones dada por las expresiones (\ref{TH_Pot_0:N1}) y (\ref{TH_Pot_B:N}). 

\begin{table}[h]
\begin{center}
\begin{tabular}{cccccc} \toprule
	Estructura & $C$ & $\eta$ & $\zeta$ & $(1-\eta-\zeta)$ & $\xi$ \\ \midrule
	sc & $-1.418649$ & $0.403981$ & $0.403981$ & $0.192037$ & $1/2$ \\
	fcc & $-1.444141$ & $0.654710$ & $0.154710$ & $0.190580$ & $3/4$ \\
	bcc & $-1.444231$ & $0.389821$ & $0.389821$ & $0.220358$ & $1/2$\\
	hcp & $-1.444083$ & $0.345284$ & $0.345284$ & $0.309433$ & $1/2$\\ \bottomrule
\end{tabular}
\end{center}
\caption{Parámetros $C$, $\eta$, $\zeta$ y $(1-\eta-\zeta)$ de la red para diferentes estructuras cristalinas heterogéneas \cite{1982ApJ...253..839J,PhysRevD.92.023008} obtenidos mediante el método de Coldwell-Horsfall y Maradudin, y fracción de iones del tipo \mychemistryg{X}{A}{Z} consideradas.}
\label{tab:lcte}
\end{table}%

La corrección a la presión introducida al tener en cuenta la red es:
\begin{equation}
	\label{eq:latticeP}
	P_L = \frac{\varepsilon_L}{3}.
\end{equation}

Luego, la densidad de energía total y las presiones, i. e., las ecuaciones de estado se modifican debido a la energía y la presión de la red, y quedan:
\begin{subequations} \label{eq:latticeEOS}
\begin{align} \label{eq:latticeET}
	E &= m_{\mathrm{X}}N_{\mathrm{X}} + m_{\mathrm{X}’}N_{\mathrm{X}’}- m\,N (x,0,B) + m^4 \chi (x,B) + \varepsilon_L , \\  \label{eq:latticePll}
	P_\parallel &=  m^4 \Phi_+(x,B) +P_L ,\\  \label{eq:latticePp}
	P_\perp & = m^4 \Phi_-(x,B) +P_L,
\end{align}
\end{subequations}
donde la función $m^4\chi(x,B)$ se corresponde con la densidad de energía del gas de electrones:
\begin{equation}
	\chi (x,B) =\chi (x)+\frac{1}{12\pi^2} \left(\frac{B}{B_c}\right)^2\!\left[\frac{\sqrt{x^2+1}}{x}-\ln\!\left(x+\sqrt{x^2+1}\right)\right];
\end{equation}
$m_{\mathrm{X}}$ ($m_{\mathrm{X}’}$) y $N_{\mathrm{X}}$ ($N_{\mathrm{X}’}$) son, respectivamente, la masa en reposo del átomo \mychemistryg{X}{A}{Z} (\mychemistryg{X$’$}{A$’$}{Z$’$}), y la densidad de los iones de tipo correspondiente. Para no contar doblemente los electrones, se sustrajo el término $m\,N(x,0,B)$. 

Usando la condición de neutralidad de carga:
\begin{equation}
	N(x,0,B) = \mathrm{Z} N_{\mathrm{X}} +\mathrm{Z}’ N_{\mathrm{X}’},
\end{equation}
se obtiene:
\begin{align} \label{eq:NX}
	N_{\mathrm{X}} &= \frac{\xi}{\mathrm{Z}\xi+(1-\xi)\mathrm{Z}’} N(x,0,B), \\
	N_{\mathrm{X}’}  &= \frac{1-\xi}{\mathrm{Z}\xi+(1-\xi)\mathrm{Z}’} N(x,0,B).
\end{align}

Si consideramos el caso homogéneo, donde solo tenemos un tipo de ión en la red cristalina, la energía de la red se reduce a:
\begin{equation} \label{eq:latticeE1}
	\varepsilon_{L} = Ce^2 [N(x,0,B)]^{4/3} Z^{2/3},
\end{equation}
y la condición de neutralidad de carga a:
\begin{equation}
	N(x,0,B) = \mathrm{Z} N_{\mathrm{X}}.
\end{equation}
Teniendo en cuenta, además, que la densidad de energían en reposo del átomo  \mychemistryg{X}{A}{Z} es:
\begin{equation}
	m_{\mathrm{X}}N_{\mathrm{X}} = \frac{A}{Z}m_N N(x,0,B) + m\,N (x,0,B),
\end{equation}
podemos reescribir las ecuaciones de estado como:
\begin{subequations} \label{eq:latticeEOS1}
\begin{align} \label{eq:latticeET1}
	E &= \frac{A}{Z}m_N N(x,0,B) + m^4 \chi (x,B) + \varepsilon_{L} , \\  \label{eq:latticePll1}
	P_\parallel &=  m^4 \Phi_+(x,B) +P_{L} ,\\  \label{eq:latticePp1}
	P_\perp & = m^4 \Phi_-(x,B) +P_{L},
\end{align}
\end{subequations}

Nótese que en este caso, la energía es la suma de la expresión obtenida cuando solo se considera el aporte de la masa en reposo de los iones (\ref{eq:Eb}) con la energía de la red (\ref{eq:latticeE}), y que esta última depende únicamente del parámetro $C$, el cual varía según la red utilizada. 

\begin{figure}[h]
   \centering
   \includegraphics[width=.8\textwidth]{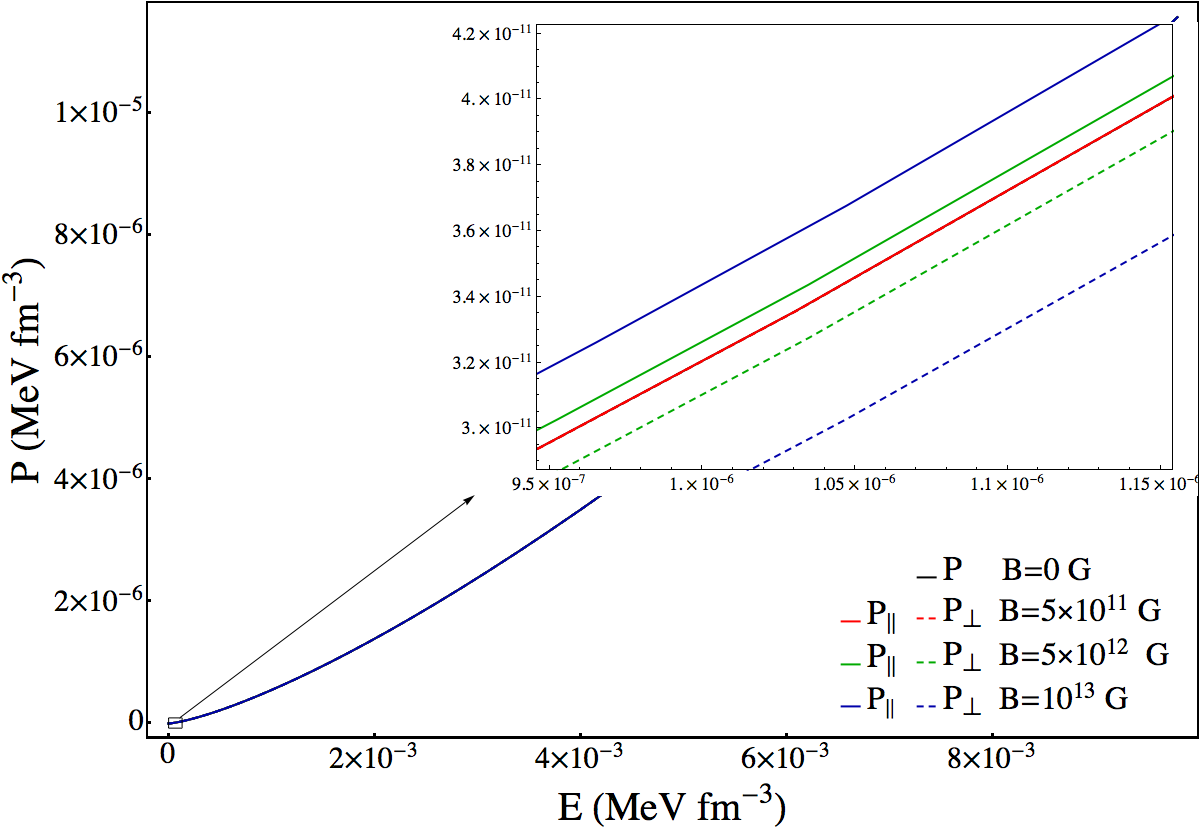} 
   \caption{Ecuaciones de estado de EBs compuestas de \mychemistry{C}{12}  ($A/Z=2$) para la red cristalina homogénea de tipo bcc.}
   \label{fig:eosglatt}
\end{figure}

\begin{figure}[h]
   \centering
   \begin{tikzpicture}
   	\node[anchor=south west,inner sep=0] at (0,0) {\includegraphics[width=.8\textwidth]{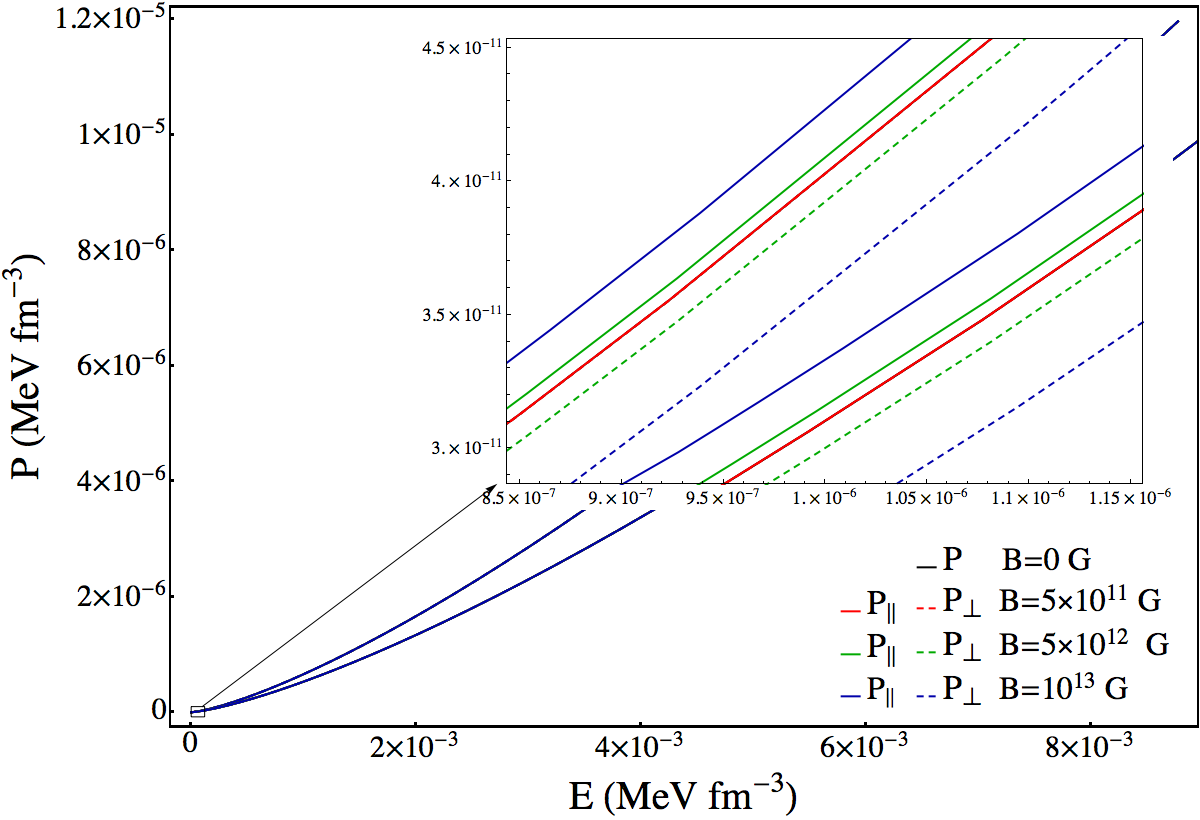}};
	\node[fill=white] at (10.2,8) {\centering \footnotesize{partículas no interactuantes}};
	\node[fill=white,text width=3.6cm] at (10.2,4.5) {\centering \footnotesize{partículas interactuantes (iones en una red bcc)}};
   \end{tikzpicture}
    \caption{Comparación entre las ecuaciones de estado representadas en las Figuras \ref{fig:eosg} y \ref{fig:eosglatt}. Nótese que al introducir la corrección las ecuaciones de estados se suavizan.}
   \label{fig:eosgCompare}
\end{figure}

En la  \Fref{fig:eosglatt} se muestran los gráficos de las ecuaciones de estado para una red bcc homogénea formada por  \mychemistry{C}{12}, que comparamos con las correspondientes al caso de partículas no interactuantes en la \Fref{fig:eosgCompare}.


\chapter{Ecuaciones de estructura}
\label{cap4}

La estructura de una estrella está determinada por la dependencia con el radio de las magnitudes termodinámicas y los coeficientes métricos. Para obtener dicha relación se emplean las ecuaciones de Einstein, que vinculan las ecuaciones de estructura y las ecuaciones de estado.

En este capítulo, se discute las soluciones de las ecuaciones de estructura en la geometría esférica usada habitualmente, y una geometría axisimétrica \cite{Paret:2015RAA,phdDaryel}, más adecuada para nuestro problema debido  a la anisotropía que introduce el campo magnético. Las EdE utilizadas son las obtenidas previamente en el Capítulo \ref{cap3}.

\section{Ecuaciones de Einstein}

La Teoría General de la Relatividad (TGR) tiene como premisa que \emph{la gravitación es la manifestación dinámica de la curvatura del espacio-tiempo}; y ha sido validada experimentalmente por los efectos que predice \cite{Misner1973grav} y no se justifican mediante la teoría de la relatividad especial \cite{2011PhDT........15U}, como las trayectorias curvas de los rayos de luz, el corrimiento hacia el rojo de un fotón bajo la influencia del campo gravitatorio, la precesión en la órbita de Mercurio, entre otros \cite{Misner1973grav}.

En la TGR, la curvatura del espacio-tiempo se debe al flujo de materia-energía. Luego, las ecuaciones de campo deben tener como fuente al tensor de energía-momento $ T_{\mu\nu}$, que es un tensor de segundo orden. Por lo tanto, la curvatura del espacio-tiempo debe expresarse por un tensor de curvatura de igual orden. Las ecuaciones de Einstein establecen este vínculo entre el contenido de materia en el espacio-tiempo y la curvatura del mismo \cite{Misner1973grav}:
\begin{equation}
G_{\mu\nu}=\kappa T_{\mu\nu},
\label{EE1}
\end{equation}
donde $x_\mu=(x_4,\vec x)$, $\kappa=8\pi \text{G}$, siendo G la constante de gravitación (ver \Fref{tab:cons} en el Apéndice \ref{appB}). El tensor de Einstein: 
\begin{equation}
	\mathrm{G}_{{\mu \nu}}=R_{{\mu \nu}}-\frac{1}{2}R g_{{\mu \nu}},
\end{equation}
está determinado por el tensor de Ricci:
\begin{equation}
R_{\mu\nu}=\Gamma^{\alpha}_{\mu\nu,\alpha}-\Gamma^{\alpha}_{\mu\alpha,\nu}+\Gamma^{\alpha}_{\mu\nu}
\Gamma^{\beta}_{\alpha\beta}-\Gamma^{\beta}_{\mu\alpha}
\Gamma^{\alpha}_{\nu\beta}, \label{TF2}
\end{equation}
y por el escalar de Ricci $R=R^{\mu}_{\,\,\,\,\mu}$, que dependen de segundas derivadas de la métrica a través de los índices de Christoffel  $\Gamma^{\alpha}_{\mu\nu}$:
\begin{equation}
\Gamma^{\alpha}_{\mu\nu}=\frac{g^{\alpha\beta}}{2}(g_{\beta\mu,\nu}+g_{\nu\beta
,\mu}-g_{\mu\nu,\beta}). \label{TF3}
\end{equation}

Los  tensores que aparecen en las ecuaciones de Einstein son simétricos, de modo que en 4 dimensiones tienen 10 componentes independientes. Dada la libertad de elección de las cuatro coordenadas del espacio-tiempo, las ecuaciones independientes se reducen a 6. Las ecuaciones de Einstein son un sistema de ecuaciones diferenciales parciales no lineales con alta complejidad, por lo que es difícil encontrar soluciones exactas.

Dadas las condiciones extremas de masas, radios y densidades que presentan los objetos compactos, las correcciones de la Teoría General de la Relatividad a las ecuaciones de Newton son importantes para una correcta descripción del equilibrio hidrostático. Particularmente, en una enana blanca la aproximación Newtoniana arroja buenos resultados, pero una correcta predicción de la masa máxima de estas estrellas debe hacerse utilizando la TGR.

En este trabajo, usaremos como fuente el tensor de energía-momento presentado en el Capítulo \ref{cap2}, teniendo en cuenta las ecuaciones de estado obtenidas en el Capítulo \ref{cap3}.

\section{Simetría  esférica: Ecuaciones de Tolman-Oppenheimer-Volkoff}

Para describir una estrella estática en equilibrio se emplea generalmente la métrica esférica:
\begin{equation}\label{MetricaSch}
    ds^{2} = -e^{2\Phi(r)}dt^{2} + e^{\Lambda(r)}dr^{2} + r^2 d\theta^{2}    +r^2 \sin^2 \theta d\phi^2
\end{equation}

Utilizando la métrica~(\ref{MetricaSch}), la ecuación~(\ref{EE1}) y la ley de conservación de la energía $(T^{\mu\nu}_{\phantom{\mu\nu};\nu}=0)$ podemos encontrar las ecuaciones de Tolman-Oppenheimer-Volkoff (TOV)~\cite{Misner1973grav,1939PhRv...55..374O}:
\begin{subequations}\label{TOV}
\begin{eqnarray}
  \frac{dM}{dr} &=& 4\pi r^2 E(r), \\
  \frac{dP}{dr} &=& -\frac{(E(r)+P(r))(M(r)+4\pi P(r)r^3)}{r^2-2rM(r)},
\end{eqnarray}
\end{subequations}
cuyas soluciones describen las configuraciones de estrellas estáticas y con simetría esférica.

\begin{figure}[h]
    \centering
    \begin{subfigure}{.47\textwidth}
        \centering
        \includegraphics[width=\textwidth]{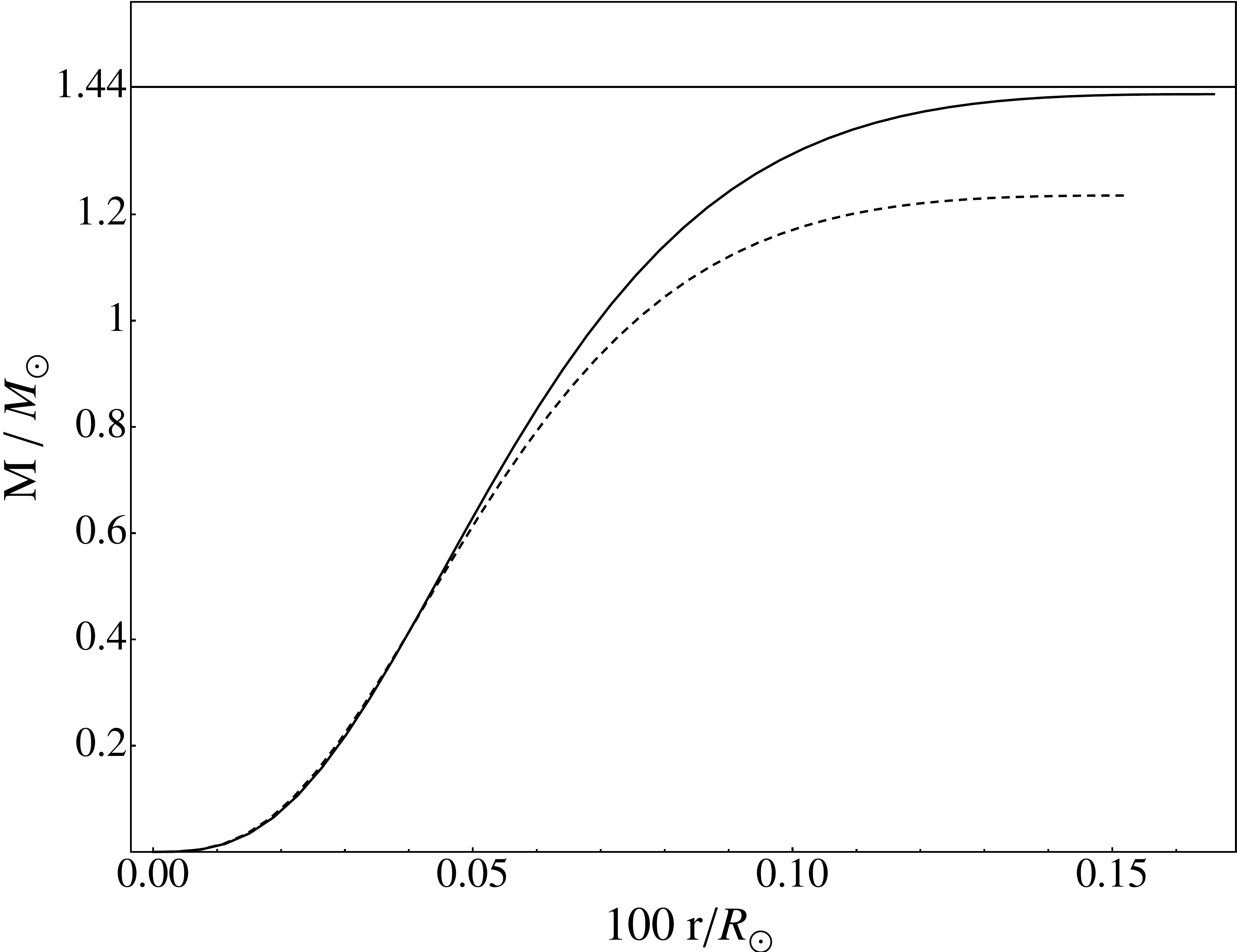}
        \label{subfig:mvsr}
    \end{subfigure}~
    \begin{subfigure}{.515\textwidth}
        \centering
        \includegraphics[width=\textwidth]{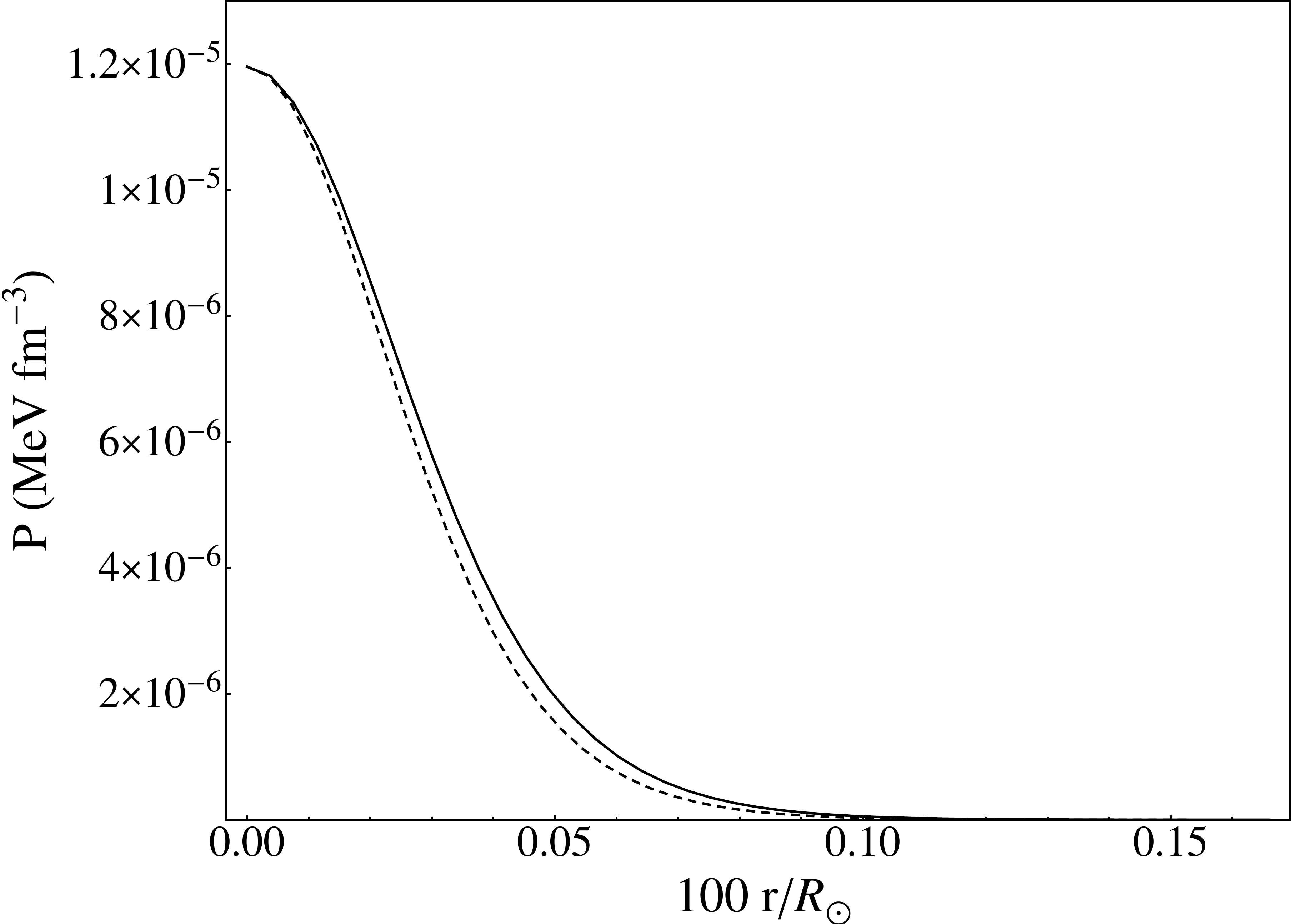}
        \label{subfig:pvsr}
    \end{subfigure}
    \caption{Funciones (a) $M(r)$ y (b) $P(r)$. Ambas se obtuvieron al resolver  las ecuaciones (\ref{TOV}) con \mbox{$E_c=E_{A/Z}(20,0,0)$} en coordenadas esféricas. La curva continua corresponde configuraciones con $A/Z=2$, y la discontinua para $A/Z=2.15$.}
    \label{fig:esfMPr}
\end{figure}

Para resolver el sistema de ecuaciones~(\ref{TOV}) utilizamos las ecuaciones de estado paramétricas halladas en el Capítulo \ref{cap3}. El radio $R$ y la masa correspondiente $M$ de la estrella se determinan imponiendo la condición de presión cero $P(R)=0$. La presión central queda fijada por la EdE, $P(0)=P_c$, y $M(0)=0$. Las funciones $M(r)$ y $P(r)$ resultantes para momentum de Fermi $x=20$ y $B=0$, por ejemplo, se comportan según la Figura \ref{fig:esfMPr}.

Dada una EdE, existe una familia única de estrellas parametrizadas por la densidad de energía central, que especifica un modelo de secuencias estelares $M=M(E_c)$, $R=R(E_c)$. Estos resultados se presentan mediante una curva donde cada punto representa una estrella de masa $M$ y radio $R$ en equilibrio hidrodinámico (diagrama masa-radio).

No todas las ramas de una secuencia M-R son estables. El problema de Sturm-Liouville correspondiente para los modos de oscilación fue tratado por primera vez por Chandrasekhar en 1964 \cite{1964ApJ...140..417C}. El criterio usual de estabilidad para las estrellas \cite{1966ApJ...145..505B} es que un modo radial se vuelve estable o inestable en cada extremo de la función $M(R)$. A medida que la densidad central aumenta, un modo estable se vuelve inestable cuando la curva gira en contra de las manecillas del reloj, y un modo inestable se hace estable donde la curva gira a favor de las manecillas del reloj.

\subsection{Discusión de los resultados numéricos}

En las ecuaciones TOV~(\ref{TOV}),  obtenidas en simetría esférica, aparece solamente una presión. En los estudios que incluye presiones anisotrópicas debido al campo magnético, se ha justificado el uso de estas ecuaciones de equilibrio hidrostático cuando el campo magnético es moderado o débil porque la diferencia entre las presiones es pequeña y, o bien se aproxima $P_{\parallel}=P_{\perp}$ o se  usa la menor de ellas \cite{2008PhRvC..77a5807G}.

En este capítulo, hemos resuelto dichas ecuaciones de estructura para las presiones paralela y perpendicular que aparecen en el Capítulo \ref{cap3}, teniendo en cuenta la presencia del campo magnético en el régimen de campo débil y  distintas composiciones químicas. Además, los iones, que interactúan entre ellos y con los electrones, se consideran distribuidos en una red cristalina. 

\begin{figure}[p]
    \centering
    \begin{subfigure}{.51\textwidth}
        \centering
        \includegraphics[width=\textwidth]{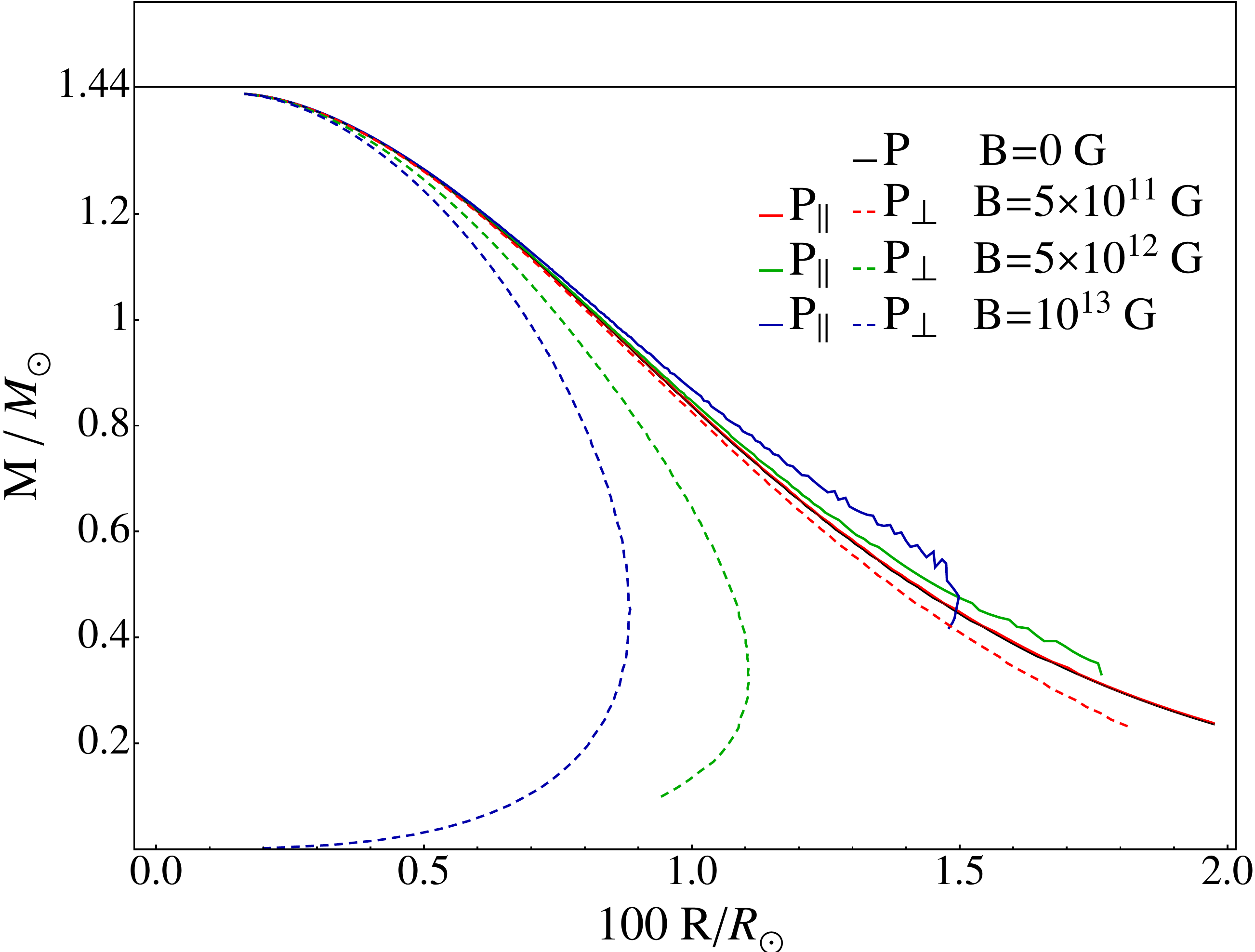}
        \caption{ \mychemistry{He}{4}, \mychemistry{C}{12}, \mychemistry{O}{16}, o \mychemistry{Mg}{24}}
        \label{subfig:esf}
    \end{subfigure}~
    \begin{subfigure}{.51\textwidth}
        \centering
        \includegraphics[width=\textwidth]{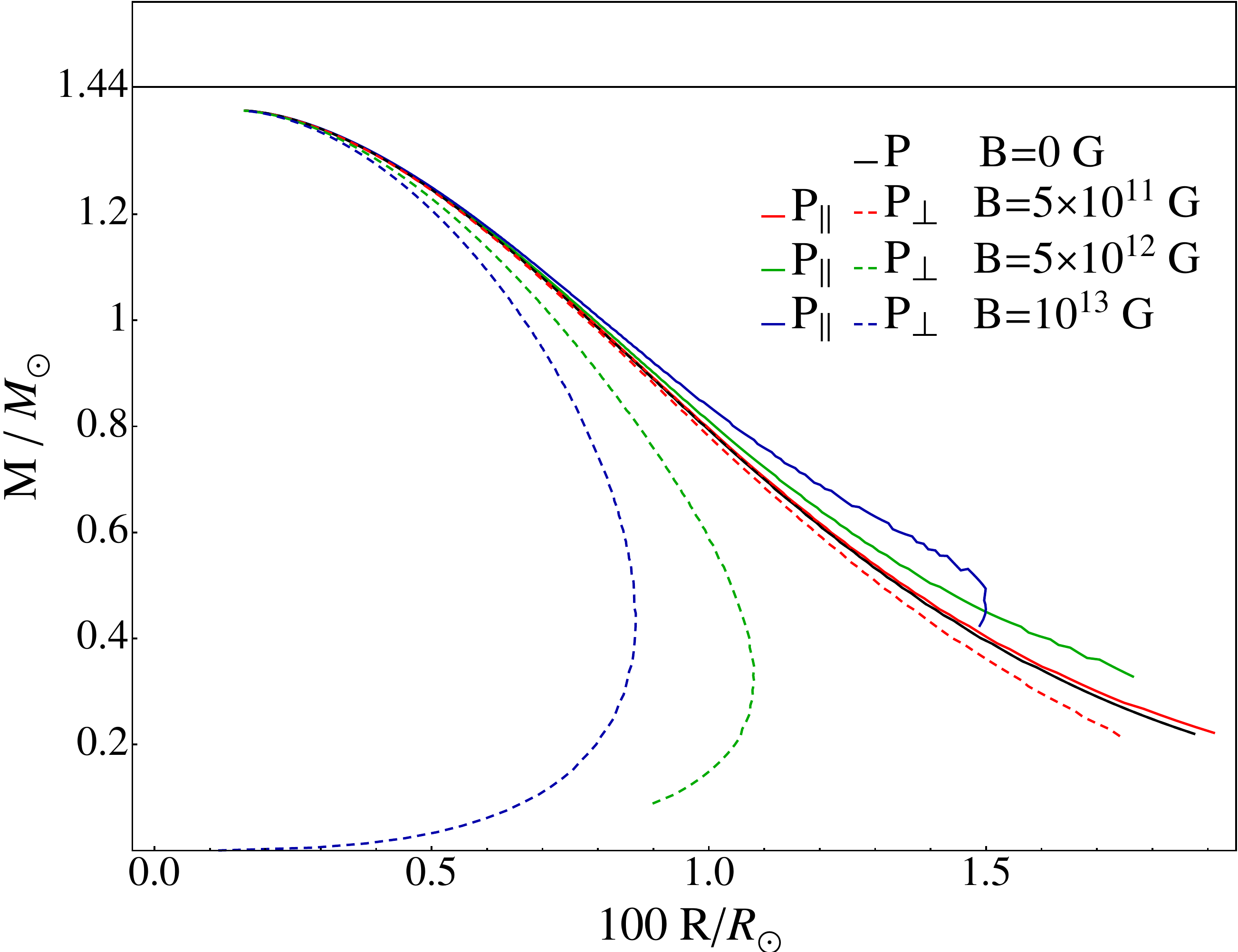}
        \caption{ \mychemistry{C}{12}}
        \label{subfig:esfint}
    \end{subfigure} \\ \vspace{12pt}
    \begin{subfigure}{.51\textwidth}
        \centering
        \includegraphics[width=\textwidth]{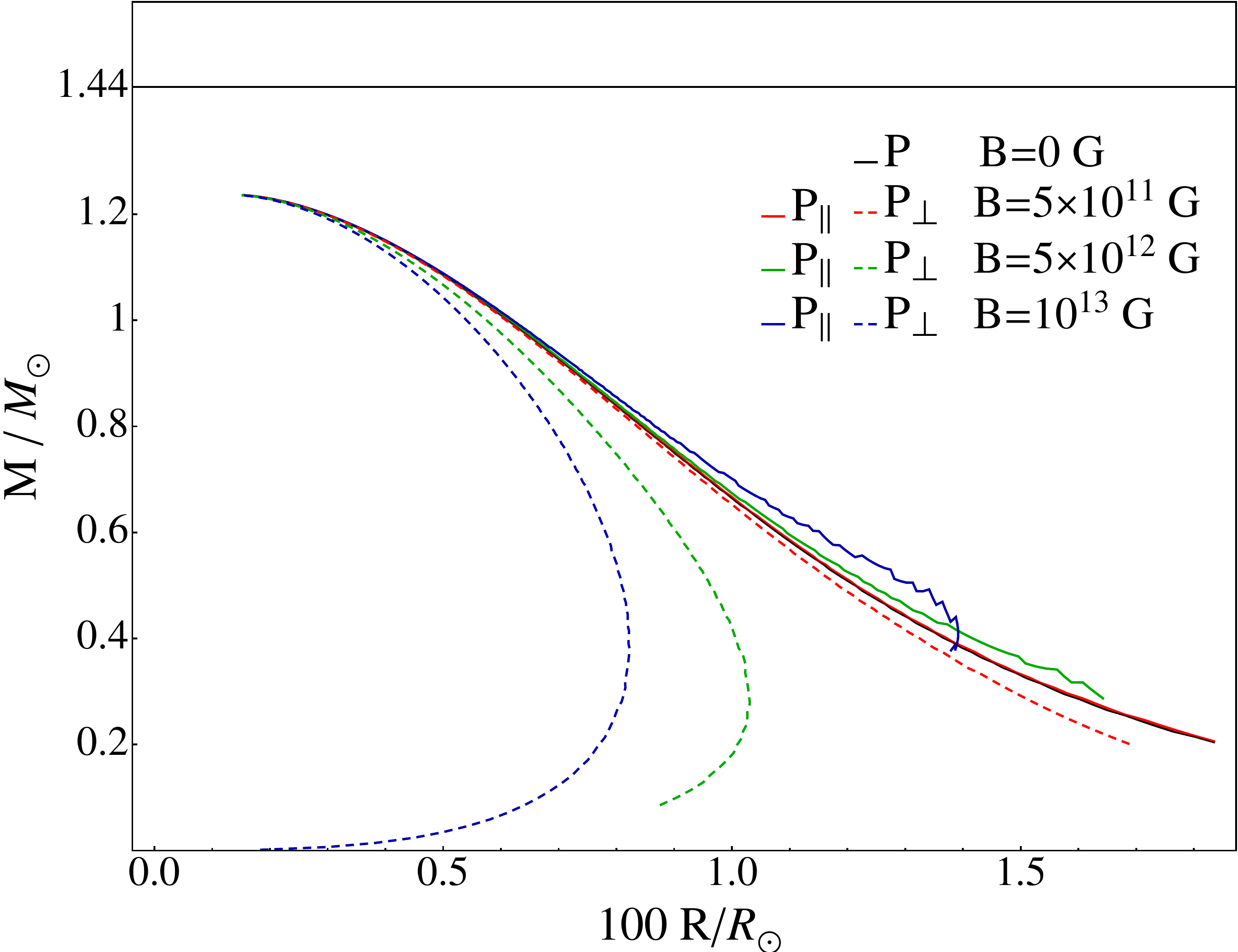}
        \caption{\mychemistry{Fe}{56}}
        \label{subfig:esfFe}
    \end{subfigure}~
    \begin{subfigure}{.51\textwidth}
        \centering
        \includegraphics[width=\textwidth]{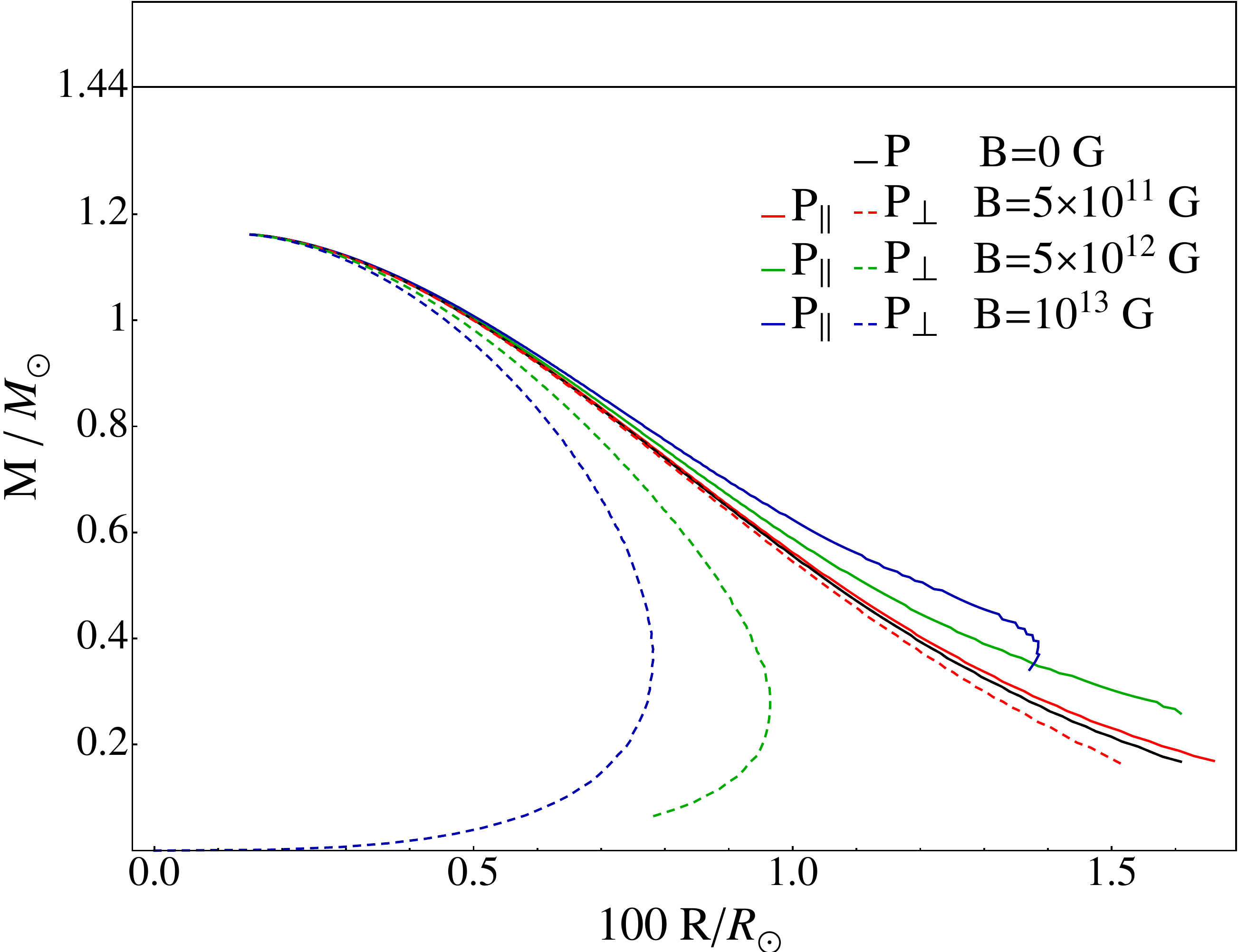}
        \caption{ \mychemistry{Fe}{56}}
        \label{subfig:esfintFe}
    \end{subfigure}
    \caption{Soluciones en simetría esférica para distintas configuraciones. En (a) y (c) se emplearon las ecuaciones de estado con componentes no interactuantes, mientras en (b) y (d) se incluye la corrección debida a la interacción de los electrones con la red, para la red de tipo bcc.}  \label{fig:esf}
\end{figure}

Nuestros resultados muestran que para una presión y campo magnético dados, con la energía como función de la composición, la masa es menor en tanto mayor sea $A/Z$ (ver diferencia entre la Figura \ref{subfig:esf} y la Figura \ref{subfig:esfFe}). Asimismo, los efectos de la corrección en las ecuaciones de estado se manifiestan en las relaciones masa-radio, ya que la masa disminuye al introducir la presión y la energía de la red (ver cambios en la Figura \ref{subfig:esfint} respecto a la Figura \ref{subfig:esf}).

Por otra parte, en la Figura \ref{subfig:esf}, notamos que para una misma composición química, a pesar de la pequeña variación entre las EdE para la presión paralela y la correspondiente presión perpendicular (\Fref{fig:eosg}), las respectivas relaciones masa-radio obtenidas son distintas, y la diferencia aumenta a medida que se incrementa el campo magnético. Este resultado ya había sido obtenido en \cite{Paret:2015RAA,phdDaryel}, donde se resolvieron las ecuaciones TOV para  presiones anisotrópicas del gas de electrones magnetizado no interactuante.

En la \Fref{fig:esf} se aprecia que las masas obtenidas a partir de las soluciones para la presión paralela y campos entre $10^{11}$ G y $10^{13}$  G (correspondientes a las obtenidas por Mathews \cite{2000ApJ...530..949S} para presión isotrópica) son mayores que las computadas en ausencia de campo. Además, no obtenemos ninguna configuración estable de enanas blancas magnéticas cuyo valor máximo de masa  sobrepase el límite de Chandrasekhar como ha sido propuesto en \cite{2013PhRvL.110g1102D}.

\begin{figure}[h]
    \centering
    \begin{subfigure}{.5\textwidth}
        \centering
        \includegraphics[width=\textwidth]{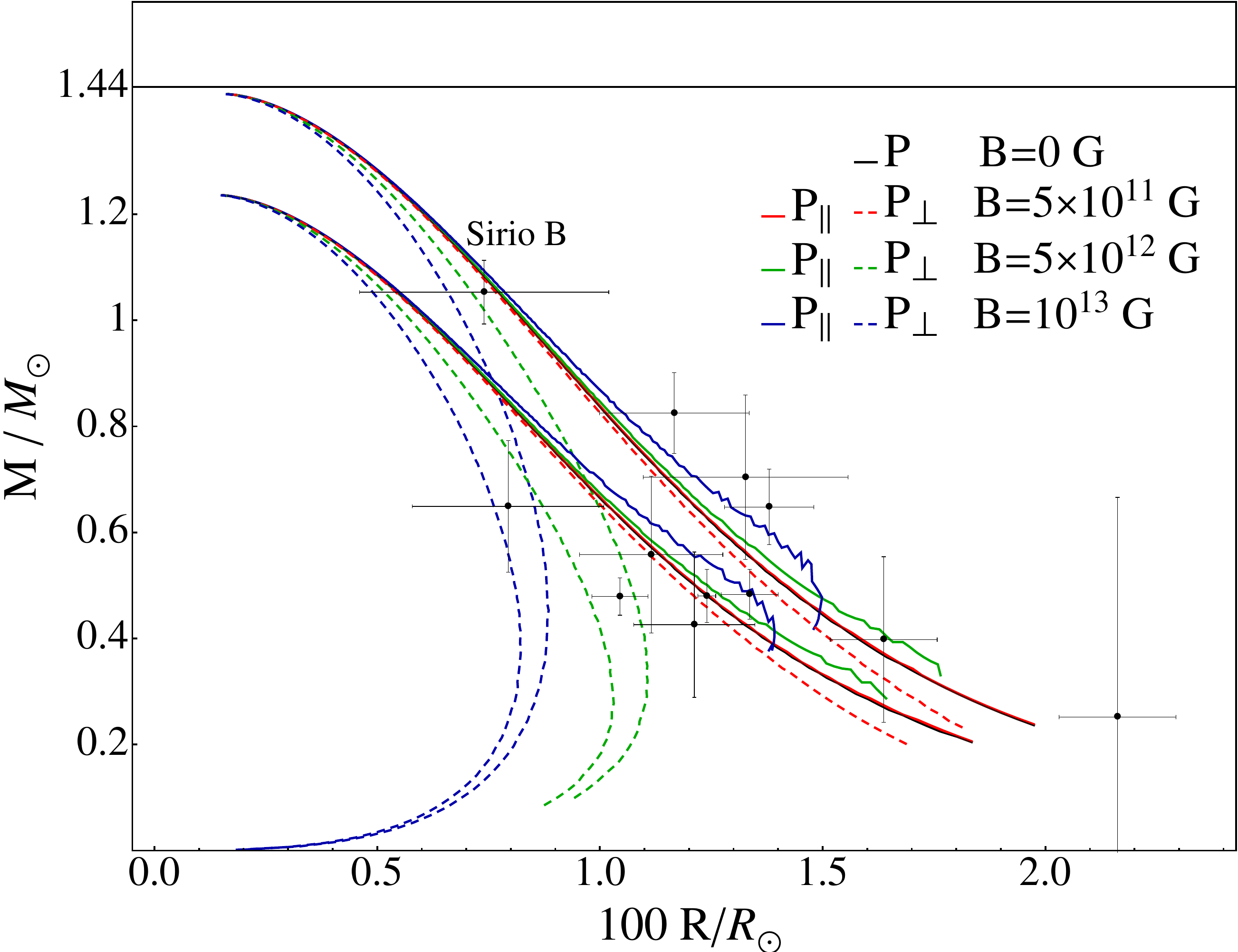}
        \label{subfig:obs}
    \end{subfigure}~
    \begin{subfigure}{.5\textwidth}
        \centering
        \includegraphics[width=\textwidth]{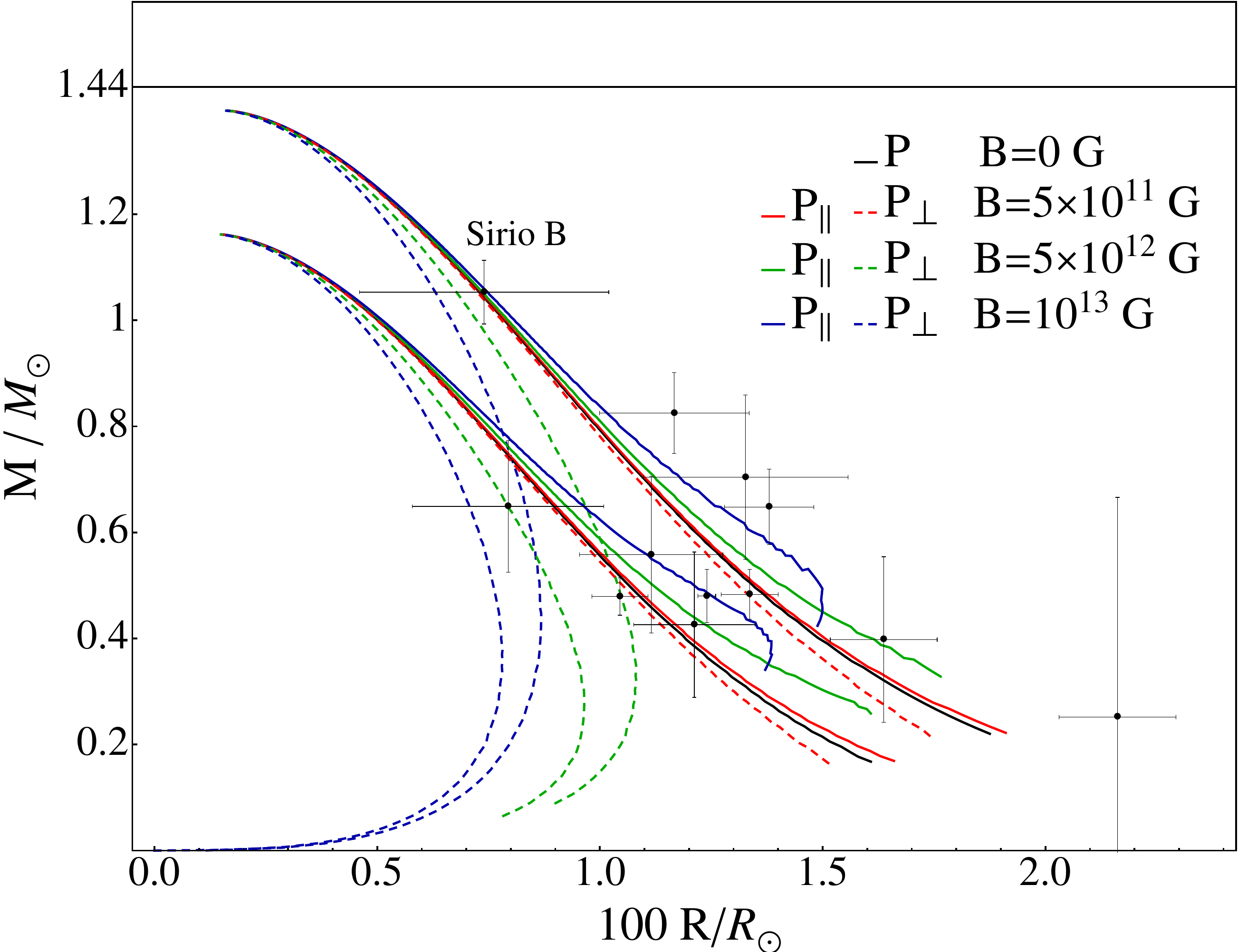}
        \label{subfig:obs1}
    \end{subfigure}
    \caption{Comparación con datos observacionales. En (a) se incluyen las soluciones para los componentes no interactuantes estudiadas en las Figuras \ref{subfig:esf} y  \ref{subfig:esfFe}; mientras en (b) se grafican las soluciones para componentes interactuantes presentadas en las Figuras \ref{subfig:esfint} y  \ref{subfig:esfintFe}.}
    \label{fig:esf1}
\end{figure}

Por otra parte, en la \Fref{fig:esf1}, se superponen con las masas y los radios observacionales de algunas enanas blancas detectadas por el satélite \emph{Hipparcos} \cite{1997A&A...325.1055V}, lo cual permite validar nuestro modelo.

Consecuentemente nuestros resultados apuntan a que es un  imperativo  emplear una métrica axisimétrica, apropiada para tomar en cuenta las ecuaciones de estado anisotrópicas que se derivan de la inclusión de la presencia del  campo magnético (ver Sección \ref{c4secCyl}) aún cuando el campo magnético sea considerado débil.

\section{Simetría cilíndrica}\label{c4secCyl}

Debido a la existencia de presiones anisotrópicas, nos propusimos estudiar la estructura de una EB en una geometría con simetría axial, que es más ``natural'' para los sistemas de electrones magnetizados. Por tanto, para obtener las ecuaciones de estructura utilizaremos la métrica:
\begin{equation}\label{cyl1}
  ds^2=-e^{2\Phi(r)} dt^2+e^{2\Lambda(r)} dr^2+r^2d\phi^2+e^{2\Psi(r)} dz^2,
\end{equation}
donde $\Phi$, $\Lambda$, y $\Psi$ son funciones de $r$ solamente.

De las ecuaciones de Einstein en unidades naturales y de la conservación de la energía $(T^{\mu\nu}_{\phantom{\mu\nu};\nu})$ obtenemos el siguiente sistema de ecuaciones diferenciales \cite{Trendafilova2011EJPh}:
\begin{subequations}\label{Diff2}
\begin{eqnarray}
P_{\perp}'&=&-\Phi'(E+P_{\perp})-\Psi'(P_{\perp}-P_{\parallel}),\\
4\pi e^{2\Lambda}(E+P_{\parallel}+2P_{\perp})&=&\Phi''+\Phi'(\Psi'+\Phi'-\Lambda')+\frac{\Phi'}{r}, \\
4\pi e^{2\Lambda}(E+P_{\parallel}-2P_{\perp})&=&-\Psi''-\Psi'(\Psi'+\Phi'-\Lambda')-\frac{\Psi'}{r}, \\
4\pi e^{2\Lambda}(P_{\parallel}-E)&=&\frac{1}{r}(\Psi'+\Phi'-\Lambda').
\end{eqnarray}
\end{subequations}
Las expresiones anteriores, junto a las EdE $E(x) =  E(x(P_{\perp}))$, y $P_{\parallel} = P_{\parallel}(x)$ forman un sistema de ecuaciones en las variables $P_{\perp}$, $P_{\parallel}$, $E$, $ \Phi$, $ \Lambda$, $\Psi$, $ \Phi’$, y $ \Psi’$.

Las condiciones de frontera empleadas deben tener en cuenta los factores del tipo $1/r$ en (\ref{Diff2}). Por tanto,  se hace una expansión en serie de $P_{\perp}$, $\Phi$, $\Psi$, y $\Lambda$ alrededor de $r=0$. También, $\Psi=\Phi=\Lambda=0$ en $r=0$ para que los coeficientes métricos correspondientes  sean igual a 1 en ese punto y $\Psi'=\Phi'=0$ para que las soluciones en el eje $z$ sean suaves. Con estas consideraciones, las condiciones de frontera en el centro son:
\begin{subequations}\label{IC1}
\begin{eqnarray}
  P_{\perp}(0) &=&P_{\perp0},  \\
  \Phi(0) &=& \frac{1}{2}(P_{\parallel0}+2P_{\perp0}+E_0)(r_0^2-2r_0), \\
  \Psi(0) &=& \frac{1}{2}(-P_{\parallel0}+2P_{\perp0}-E_0)(r_0^2-2r_0), \\
  \Phi'(0) &=& \Psi'(0)= \Lambda(0)=0.
\end{eqnarray}
\end{subequations}

 Además, se impone la condición $P_{\perp}(R_{\perp})=0$, para determinar el radio de la estrella en la dirección ecuatorial (perpendicular).

Por hipótesis, en este modelo, todas las variables dependen solamente de la coordenada radial (en el plano perpendicular al campo magnético). Luego, no es posible calcular la masa total de la estrella como se hizo para el caso de simetría esférica. Por tanto calcularemos la generalización para la masa dada por Tolman~\cite{Tolman1934rtc}
\begin{equation}\label{tolman1}
  M_T=\int\sqrt{-g}(T^0_0-T^1_1-T^2_2-T^3_3)dV
\end{equation}
para la métrica cilíndrica (\ref{cyl1}) no podemos calcular la masa, sino, la masa por unidad de longitud $(M_T/R_{\parallel})$ \cite{Paret:2014sba}:
\begin{equation}\label{tlman3}
   \frac{M_T}{R_{\parallel}}=4\pi\int_0^{R_{\perp}} re^{\Phi(r)+\Psi(r)+\Lambda(r)}(E-2P_{\perp}-P_{\parallel})dr
\end{equation}

\subsection{Discusión de los resultados numéricos}

Para la métrica cilíndrica que presentamos en esta Sección, resolvimos las ecuaciones de estructura formadas por (\ref{Diff2}) y (\ref{tlman3}), con las ecuaciones de estado del Capítulo \ref{cap3}. Al igual que en el caso de simetría esférica, hemos considerado la presencia del campo magnético débil y  las composiciones químicas ya discutidas.

\begin{figure}[h] 
    \centering
    \begin{subfigure}{.51\textwidth}
        \centering
        \includegraphics[width=\textwidth]{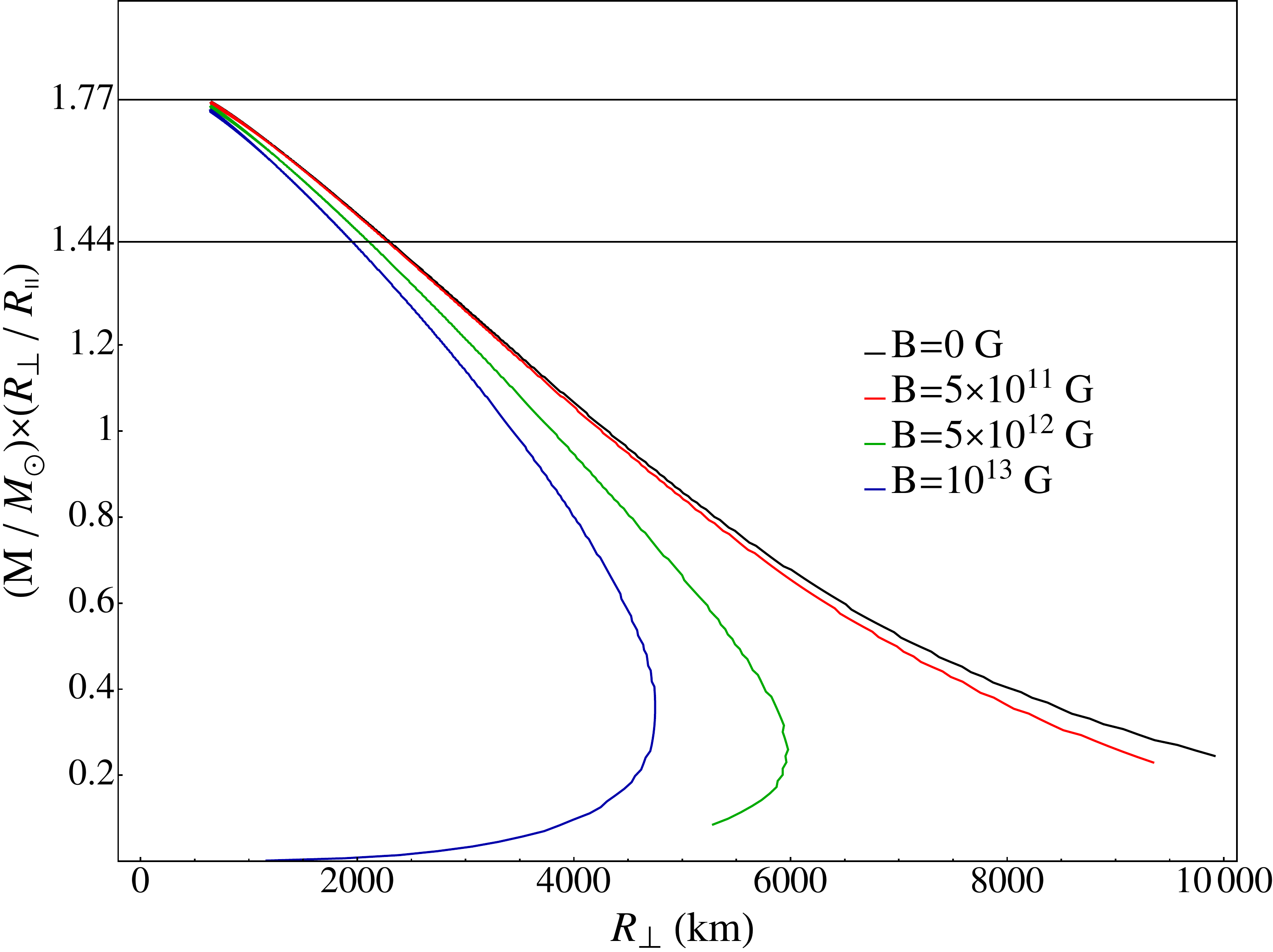}
        \caption{ \mychemistry{He}{4}, \mychemistry{C}{12}, \mychemistry{O}{16}, o \mychemistry{Mg}{24}}
        \label{subfig:cyl}
    \end{subfigure}~
    \begin{subfigure}{.51\textwidth}
        \centering
        \includegraphics[width=\textwidth]{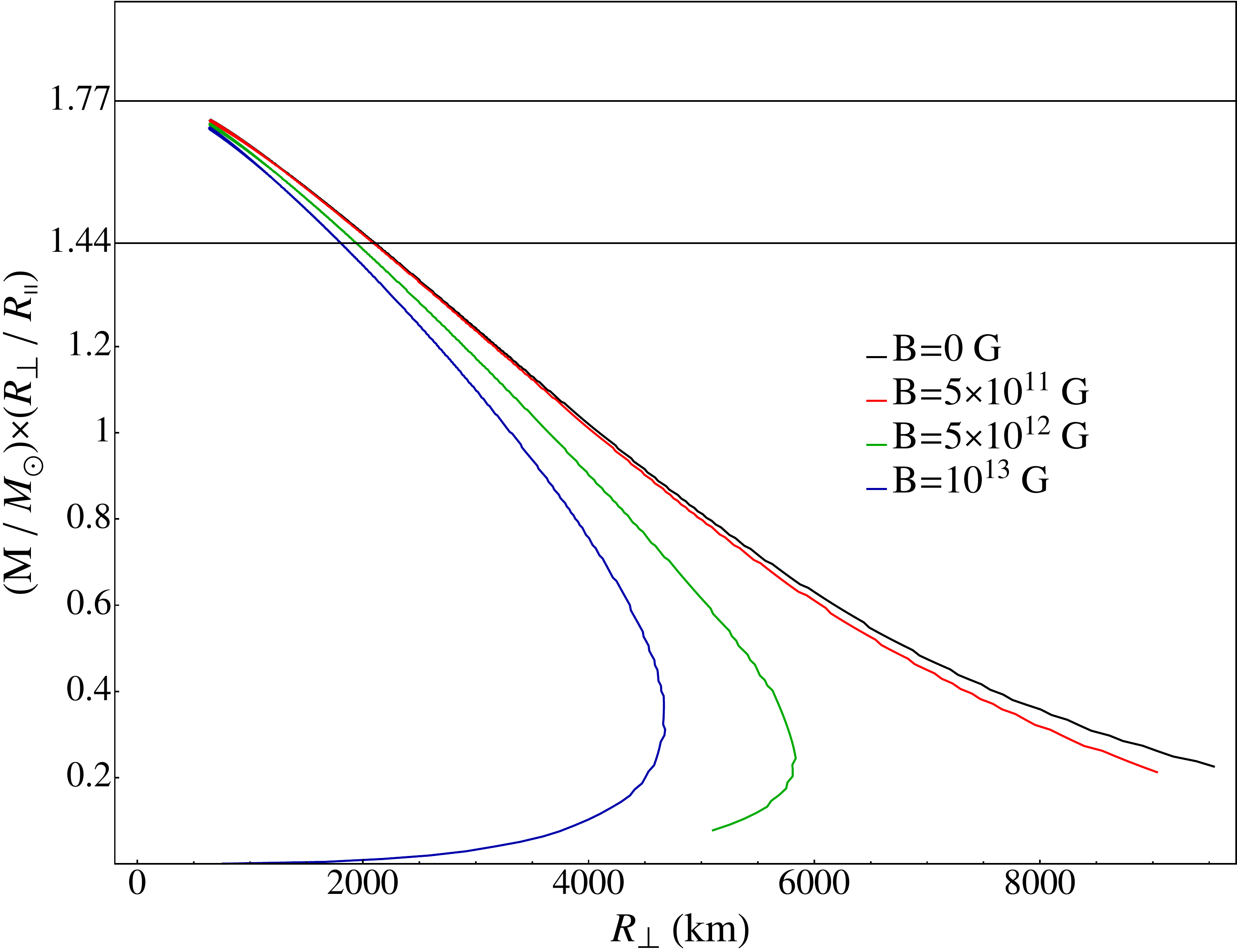}
        \caption{ \mychemistry{C}{12}}
        \label{subfig:cylTC}
    \end{subfigure} \\ \vspace{12pt}
    \begin{subfigure}{.51\textwidth}
        \centering
        \includegraphics[width=\textwidth]{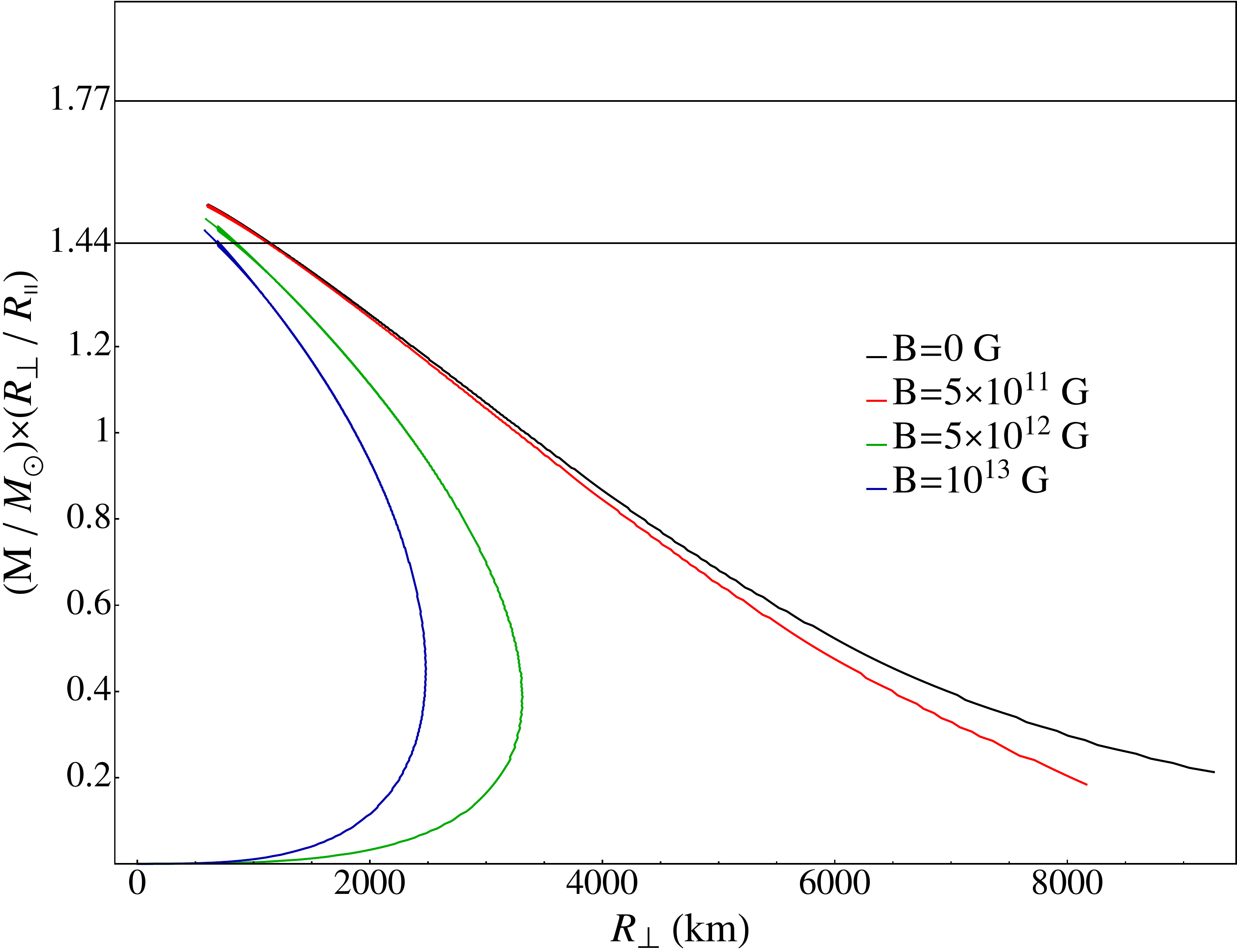}
        \caption{\mychemistry{Fe}{56}}
        \label{subfig:cylFe}
    \end{subfigure}~
    \begin{subfigure}{.51\textwidth}
        \centering
        \includegraphics[width=\textwidth]{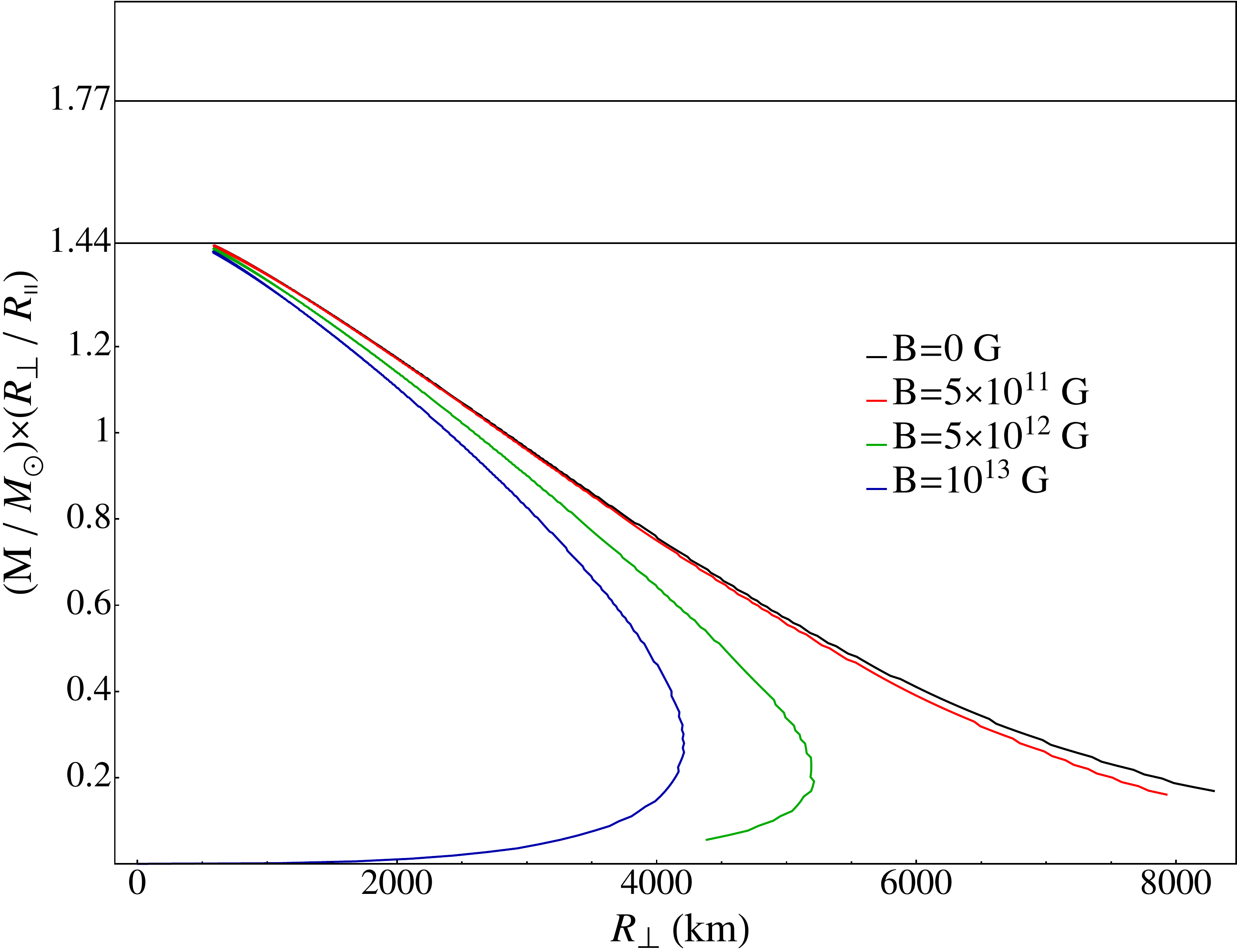}
        \caption{ \mychemistry{Fe}{56}}
        \label{subfig:cylTCFe}
    \end{subfigure}
    \caption{Soluciones en cilíndricas para distintas configuraciones. En (a) y (c) se emplearon las ecuaciones de estado con componentes no interactuantes, mientras en (b) y (d) se incluye la corrección debida a la interacción de los electrones con la red, para la red de tipo bcc.}    \label{fig:cyl}
\end{figure}

En este caso, las soluciones numéricas graficadas en la  \Fref{fig:cyl}, muestran que en la magnitud $(M_T/M_\odot)\times(R_\perp/R_{\parallel})$ se mantiene el efecto de disminución que se obtenía para la masa al variar la composición química e introducir la corrección debida a las interacciones. 

Sin embargo, a pesar de obtener un valor máximo para $(M_T/M_\odot)\times(R_\perp/R_{\parallel})$, debemos aclarar que esta no es la masa total de la estrella o masa de Schwarzchild, como se discute en  \cite{Paret:2015RAA,phdDaryel}.

\chapter*{Conclusiones}
\addcontentsline{toc}{chapter}{Conclusiones}

El trabajo presentado en esta tesis ha consistido en estudiar tanto las ecuaciones de estado como las de estructura de las enanas blancas magnetizadas, analizando los efectos de la anisotropía producida por el campo magnético, que da lugar a la presencia de dos presiones: una paralela al campo y otra perpendicular a él. Además, se tiene en cuenta que los iones no están uniformemente distribuidos y se encuentran formando una red cristalina, cuya estructura depende de la composición química de la estrella en cuestión. Por tanto, se ha incluido la corrección Coulombiana debida a la interacción de los electrones con esa red cristalina para varios tipos de redes y distintas composiciones químicas.

Podemos destacar los siguientes resultados relativos a las ecuaciones de estado:
\begin{itemize}
\item Las observaciones astrofísicas para enanas blancas magnetizadas, reportan campos magnéticos superficiales entre $10^3$ G y $10^9$ G, mientras que en el interior los  campos magnéticos  máximos permitidos son de $10^{12}$ G. Esto justifica suponer campos magnéticos en el límite de campo débil ($B<B_c=4.41\times 10^{13}$ G), lo que nos ha permitido obtener todas las magnitudes como una corrección dependiente del campo magnético respecto a las correspondientes expresiones no magnetizadas.

\item Las ecuaciones de estado que hemos obtenido para estudiar las enanas blancas magnetizadas son más realistas que las empleadas en trabajos anteriores del grupo \cite{phdDaryel}. Incluyen la presencia del campo magnético y la interacción Coulombiana entre los electrones y los iones distribuidos en una red cristalina, cuya estructura depende de la composición química de la estrella. Las expresiones están escritas de forma general, para redes heterogéneas y homogéneas, que ofrecen la posibilidad de considerar distintas composiciones químicas. Esto permite una gran libertad para estudiar de los efectos de la corrección, que esencialmente disminuye la energía y la presión del sistema.

\item Para obtener las expresiones de las energías y las presiones se siguieron los métodos de la teoría cuántica de campos a temperatura finita, lo cual garantiza que podamos generalizar nuestros resultados, a pesar de haber tomado el límite degenerado.
\end{itemize}

Con las  ecuaciones de estado así obtenidas (anisotrópicas y considerando la interacción de los electrones con la red),  se han resuelto las ecuaciones de equilibrio hidrostático en simetría esférica: las llamadas ecuaciones de Tolman-Oppenheimer-Volkoff (TOV). Igualmente, se ha determinado las soluciones de las ecuaciones de estructura en simetría cilíndrica que presentamos, más convenientes al tomar en cuenta la anisotropía de las presiones.

Entre  los resultados relacionados con las ecuaciones de estructura sobresalen los que siguen:
\begin{itemize}
\item Se resolvieron las ecuaciones TOV para las presiones anisotrópicas, obteniendo diferentes configuraciones estables de masa y radio cuando es considerada una u otra presión. Esto confirmó que aún en el límite de campo débil, aplicable para EBs magnetizadas, es necesario recurrir a una simetría axisimétrica cuando existen presiones anisotrópicas.

\item Se  reprodujeron los resultados isotrópicos obtenidos en \cite{2000ApJ...530..949S}, correspondientes con nuestra presión paralela.

\item Como era de esperar, en el régimen de campo débil, los valores máximos de masas para las configuraciones estables encontradas son siempre menores que la masa de Chandrasekhar $M_{Ch}=1.44M_\odot$.

\item En simetría cilíndrica, como consecuencia de haber supuesto que los coeficientes métricos solo dependen de $r$ y no de las otras coordenadas espaciales ($\theta$, $z$) \cite{Paret:2015RAA,phdDaryel}, no se encuentran soluciones para los valores totales de las masas por lo que no podemos obtener los usuales gráficos  de masas y radios. Sin embargo, podemos obtener configuraciones de  $(M_T/M_\odot)\times(R_\perp/R_{\parallel})$ contra $R_{\perp}$ con valores máximos.

\item Los resultados numéricos al introducir redes homogéneas, con un solo tipo de ión, formadas por los elementos  \mychemistry{He}{4}, \mychemistry{C}{12}, \mychemistry{O}{16}, o \mychemistry{Mg}{24}, que tienen $A/Z=2$; y \mychemistry{Fe}{56}, con $A/Z=2.15$ muestran que según aumenta $A/Z$ disminuye la masa (o la magnitud $(M_T/M_\odot)\times(R_\perp/R_{\parallel})$ en cilíndricas).
\end{itemize}


\chapter*{Recomendaciones}
\addcontentsline{toc}{chapter}{Recomendaciones}

Para dar continuidad a este trabajo, sugerimos:

\begin{itemize}
\item Complementar el estudio de las ecuaciones de estado teniendo en cuenta otras  composiciones químicas y otras redes cristalinas. 

\item Considerar los efectos de la temperatura, que si bien no afectarían mucho los observables astrofísicos masas y radios, podrían tener implicaciones en los fenómenos de transporte de energía que se dan en las enanas blancas, como la emisión de radiación (luminosidad) y la emisión de neutrinos. 

También, pudiera estudiarse la relación entre la intensidad del campo magnético de la estrella y su temperatura \cite{nature13836Cooling}.

\item Obtener ecuaciones de estructura en simetría cilíndrica suponiendo que los coeficientes métricos dependen de todas las coordenadas espaciales y no únicamente de $r$, de manera que la solución devuelva la masa total o la masa de Schwarzchild del objeto compacto.
\end{itemize}
    
    
    \appendix
    \addcontentsline{toc}{chapter}{Apéndices}

\chapter{Unidades y constantes físicas usadas}
\label{appB}

En la tesis, se emplean las unidades naturales (UN) cuando se tratan aspectos cuánticos. En este sistema:
\begin{equation}
\hbar=c=1, \qquad [\text{longitud}]=[\text{tiempo}]=[\text{masa}]^{-1}=[\text{energía}]^{-1}.
\end{equation}

Los cálculos y los resultados numéricos se realizaron en unidades nucleares, donde:
\begin{equation}
[\text{longitud}]=\text{fm}=10^{-15} \text{m}, \quad [\text{tiempo}]=\text{fm}/{c}=3\times10^{-24} \text{ s}, \quad [\text{masa}]=[\text{energía}]=\text{MeV}.
\end{equation}

En las expresiones de las propiedades termodinámicas que se obtuvieron en los Capítulos \ref{cap2} y \ref{cap3}, para convertir de unidades naturales a las nucleares, se divide por $(\hbar c)^3$, con $\hbar c=197.327\text{ MeV fm}$.

Por otra parte, la unidad del campo magnético usada es el Gauss (1 G$ = 10^{-4}$ T), correspondiente al sistema CGS. 

\begin{table}[ht]
\begin{tabular}{llll@{}}
  \toprule
  Magnitud Física (Símbolo) & \qquad SI & \qquad CGS & \quad UN  \\ \midrule
  Velocidad de la luz ($c$) &  $2.998\!\times\!10^{8}$  m s$^{-1}$ & $2.998\! \times\! 10^{10}$ cm s$^{-1}$  & 1  \\ 
  Carga eléctrica del electrón ($e$) & $1.602\! \times \!10^{-19}$ C  & $4.803\!\times\! 10^{-10}$ erg$^{1/2}$cm$^{1/2}$& 0.0854  \\ 
  Constante de Dirac ($\hbar$) & $1.054\! \times\! 10^{-34}$ J s   & $1.054\! \times \!10^{-27}$ erg s  & 1   \\ 
  Masa en reposo del electrón ($m$) & $9.109\! \times\! 10^{-31}$ kg  & $9.109 \!\times \!10^{-28}$ g  & 0.511 MeV \\ 
  Constante de gravitación (G) & $6.674 \!\times \!10^{-11}$ Nm$^{2}$kg$^{-2}$  & $6.674 \!\times\! 10^{-8}$ cm$^{-3}$g$^{-1}$s$^{-2}$ \\ \bottomrule
\end{tabular}
\caption{Principales constantes usadas expresadas en los sistemas SI, CGS y UN.}\label{tab:cons}
\end{table}

\begin{table}[ht]
\centering
\begin{tabular}{cccc}
  \toprule
  Magnitud Física &  Símbolo & Valor & Unidades (SI) \\ \midrule
  Masa & $M_{\odot}$ & $1.99 \times 10^{30}$ & kg \\ 
  Radio & $R_{\odot}$ & $6.96 \times 10^{5}$ & km \\ 
  Temperatura efectiva & $T^e_{\odot}$ & $5.778\times 10^{3}$ & K \\ 
  Densidad & $\rho_{\odot}$ & $1.410\times 10^{3}$ & kg m$^{-3}$ \\ 
  Luminosidad & $L_{\odot}$ & $3.846 \times 10^{26}$ & W \\ \bottomrule
\end{tabular}
\caption{Principales parámetros del Sol.}\label{tab:Sol}
\end{table}

\chapter{Contribución estadística del potencial termodinámico en el límite de campo magnético débil}
\label{appA}

Para obtener la contribución estadística del potencial termodinámico en el régimen de campo magnético débil, emplearemos la fórmula de Euler-MacLaurin (\ref{EMcL}). 

Si se aproxima hasta la segunda potencia en $(eB)$, teniendo en cuenta el número de Bernoulli $B_2=1/6$,
la función:
\begin{equation}
	\label{EMcL_app}
	f(2eBl)=\left(\mu - \sqrt{p_3^2+m^2+2eBl}\right)\Theta \left(\mu - \sqrt{p_3^2+m^2+2eBl}\right),
\end{equation}
tal que $f(\infty)=0$ y $f'(\infty)=0$, se obtiene:
\begin{equation}
	\label{EMcL1_app}
	eB \sum_{l=0}^\infty (2-\delta_{l0}) f(2eBl) \approx \int_0^\infty (2eB) f(2eBl)dl - 2 (eB)^{2} B_{2} f’(0).
\end{equation}

Por tanto, el potencial (\ref{TH_Pot_B:deg}) queda:
\begin{equation}
	\label{TH_Pot_B:st1_app}
	\Omega_{\text{est}} (\mu, 0, B) = - \frac{1}{4\pi^2 } \int_{-\infty}^\infty dp_3\left[\int_0^\infty (2eB) f(2eBl)dl - 2 (eB)^{2} B_{2} f’(0)\right].
\end{equation}

Si se define:
\begin{equation}
	\label{eq:I_app}
	I =- \frac{1}{4\pi^2 } \int_{-\infty}^\infty \int_0^\infty (2eB) f(2eBl)dl,
\end{equation}
y y se toma el límite clásico \cite{1968PhRv..173.1210C}:
\begin{equation}
	p_\perp^2= 2eBl=p_1^2+p_2^2, \qquad p_\perp dp_\perp =eBdl,
\end{equation}
obtenemos:
\begin{align}  \label{eq:Ip_app}
	I &= \int_{-\infty}^\infty  dp_3\int_0^\infty p_\perp dp_\perp \left(\mu - \sqrt{p^2+m^2}\right)\Theta \left(\mu - \sqrt{p^2+m^2}\right)\\  
	  &= - \frac{1}{4\pi^3} \int d^3\vec p \left(\mu - \sqrt{p^2+m^2}\right)\Theta \left(\mu - \sqrt{p^2+m^2}\right).
\end{align}

Nótese que $I$ es el potencial a campo cero (\ref{TH_Pot_0}). Para resolver la integral planteada, hacemos un cambio de coordenadas. Queda:
\begin{align}
 	\label{TH_Pot_0:1}
	\Omega (\mu, 0, 0) &= - \frac{1}{\pi^2} \int_0^{\infty} p^2 \left(\mu - \sqrt{p^2+m^2}\right)\Theta \left(\mu - \sqrt{p^2+m^2}\right) dp, \\
	&= - \frac{1}{\pi^2} \int_0^{\sqrt{\mu^2-m^2}} p^2 \left(\mu - \sqrt{p^2+m^2}\right) dp,
\end{align}

resultando:
\begin{equation}
	\label{TH_Pot_0:2_app}
	\Omega (\mu, 0, 0) = - \frac{m^4}{4\pi^2} \left[\frac{\mu \sqrt{\mu^2-m^2}}{3 m^2}\left(\frac{\mu^2}{m^2}-\frac{5}{2}\right) + \frac{1}{2} \ln\left(\frac{\mu+\sqrt{\mu^2-m^2}}{m}\right)\right].
\end{equation}

Sustituyendo (\ref{eq:Ip_app}) en (\ref{TH_Pot_B:st1_app}):
\begin{equation}
	\label{TH_Pot_B:st2_app}
	\Omega_{\text{est}}  (\mu, 0, B)  = \Omega  (\mu, 0, 0) +\frac{ (eB)^{2} B_{2}}{2\pi^2}\int_{-\infty}^\infty  dp_3 f’(0).
\end{equation}

Por otra parte, teniendo en cuenta:
\begin{equation}\label{eq:fder_app}
	f'(0) = \frac{1}{2} \left\{\left[1-\frac{\mu}{\sqrt{p_3^2+m^2}}\right] \delta \left(\mu-\sqrt{p_3^2+m^2}\right)-\frac{\Theta\left(\mu-\sqrt{p_3^2+m^2}\right)}{\sqrt{p_3^2+m^2}}\right\},
\end{equation}
\begin{equation}\label{eq:J_app}
	J=\int_{-\infty}^\infty  dp_3 f'(0) = - \ln\left(\frac{\mu+\sqrt{\mu^2-m^2}}{m}\right),
\end{equation}
de donde el término asociado al campo, es:
\begin{equation}
	\label{TH_Pot_B:st3_app}
	\Omega_B  = - \frac{m^4}{12\pi^2}\left[\frac{B}{B_c}\right]^2\ln\left(\frac{\mu+\sqrt{\mu^2-m^2}}{m}\right).
\end{equation}

Finalmente, podemos escribir la parte estadística del potencial termodinámico del gas degenerado de electrones en presencia de campo magnético como:
\begin{equation}
	\label{TH_Pot_B:st4_app}
	\Omega_{\text{est}}(\mu, 0, B)   = \Omega(\mu, 0, 0)   + \Omega_B.
\end{equation} 

  \backmatter

\end{document}